\documentclass[%
aps,
superscriptaddress,nofootinbib,
reprint,
]{revtex4-2}

\usepackage{xcolor}
\usepackage{graphicx}
\graphicspath{{./fig/}}
\usepackage{amsmath,amssymb,amsfonts,mathrsfs,bm}
\usepackage[%
colorlinks=true,
linkcolor=blue,
citecolor=blue,
]{hyperref}

\begin{document}
	
\title{The power of SKA to Constrain cosmological gravitational-wave backgrounds below the astrophysical foreground noise}

\author{Chengjie Fu}
\email[]{fucj@ahnu.edu.cn}
\affiliation{Department of Physics, Anhui Normal University, Wuhu, Anhui 241002, China}

\author{Jing Liu}
\email[]{corresponding author: liujing@ucas.ac.cn}
\affiliation{International Centre for Theoretical Physics Asia-Pacific, University of Chinese Academy of Sciences, Beijing 100190, China}
\affiliation{Taiji Laboratory for Gravitational Wave Universe (Beijing/Hangzhou), University of Chinese Academy of Sciences, Beijing 100049, China} 
	
\begin{abstract}
The inspirals of supermaissive black hole binaries provide a convinced gravitational wave background in the nHz band, serving as the fiducial model of the recent gravitational wave signal reported by the PTA experiments.
The uncertainties of the number of binaries contributing to each frequency bin introduce a foreground noise in the nHz and $\mu$Hz bands against the observation of the underlying gravitational wave backgrounds of the cosmological origin. 
In this work, we investigate a new method to constrain the cosmological gravitational wave strength under the astrophysical foreground.  The energy density fluctuations from cosmological gravitational-wave sources can generally trigger the formation of compact subhalos of dark matter, and the upcoming Square Kilometer Array has the ability to constrain the abundance of the subhalos at the $\mathcal{O}(1)$ level. The cosmological gravitational wave energy spectra from various sources are expected to be constrained several orders of magnitude below the astrophysical foreground, providing more strict constraints on the parameter spaces of corresponding new physics models.
\end{abstract}

\maketitle

\emph{Introduction}.
The stochastic gravitational wave backgrounds (SGWBs) represent one of the most promising observational tools for delving into the history of the early Universe and uncovering new fundamental physics \cite{Caprini:2015tfa,Kuroyanagi:2018csn,Caprini:2018mtu,Christensen:2018iqi,Renzini:2022alw}.  Recently, the Pulsar Timing Array (PTA) collaborations, including NANOGrav \cite{NANOGrav:2023gor,NANOGrav:2023hde}, EPTA \cite{EPTA:2023fyk,EPTA:2023sfo,EPTA:2023xxk}, PPTA \cite{Reardon:2023gzh}, and CPTA \cite{Xu:2023wog}, have collectively identified the first convincing evidence for a SGWB signal at frequencies in the nHz range, which have been interpreted as GWs from SMBHBs as their fiducial model~\cite{NANOGrav:2023hfp,Ellis:2023dgf,Shen:2023pan,Ghoshal:2023fhh,Bi:2023tib,Zhang:2023lzt}. 
Nonetheless, a range of potential cosmological interpretations rooted in new physics have been extensively proposed, including first-order phase transitions (FOPTs) \cite{Fujikura:2023lkn,Addazi:2023jvg,Megias:2023kiy,Xiao:2023dbb,Li:2023bxy,Ahmadvand:2023lpp,Gouttenoire:2023bqy}, domain walls \cite{Blasi:2023sej,Barman:2023fad,Babichev:2023pbf,Gelmini:2023kvo,Zhang:2023nrs}, cosmic strings \cite{Ellis:2023tsl,Kitajima:2023vre,Wang:2023len,Lazarides:2023ksx,Eichhorn:2023gat,Yamada:2023thl}, and primordial perturbations \cite{Franciolini:2023pbf,Vagnozzi:2023lwo,Inomata:2023zup,Cai:2023dls,Liu:2023ymk,Abe:2023yrw,Yi:2023mbm,Firouzjahi:2023lzg,Salvio:2023ynn,You:2023rmn,Ye:2023xyr,HosseiniMansoori:2023mqh,Balaji:2023ehk,Jin:2023wri,Das:2023nmm,Ben-Dayan:2023lwd,Liu:2023pau,Yi:2023tdk,Frosina:2023nxu,Bhaumik:2023wmw,Yuan:2023ofl,Gorji:2023sil}. 
Given the current level of observational precision, it remains uncertain whether the signal is of purely astrophysical origin, or results from combinations of astrophysical and cosmological sources~\cite{NANOGrav:2023hvm,Ellis:2023oxs}, leaving to be solved by the future advance of observatory methods. 

The information of new physics at around the MeV and GeV scales is reflected in GWs of the nHz and $\mu$Hz bands though these violent physical processes. 
Future observational projects with advanced precision have the ability to observe fainter  cosmological SGWBs predicted by a wider variety of models and larger parameter spaces. However, the existence of the astrophysical GW sources also results in a foreground noise against the precise detection of the other SGWBs~\cite{Pan:2019uyn,Biscoveanu:2020gds,Sharma:2020btq,Li:2024iua,Pan:2023naq,Lewicki:2021kmu}. Since the number of SMBHBs that contributing to each frequency bin of the order of $10$ in the nHz/$\mu$Hz bands, the statistic uncertainties introduce an intrinsic noise which largely reduce the ability of constraining cosmological sources and corresponding parameter spaces of new physics for future nHz/$\mu$Hz GW detectors. not applicable to the nHz/$\mu$Hz bands.
In this work, we consider an alternative effect of the cosmological sources, namely, triggering the formation of substructures of dark matter, and find that these subhalos can be observed within the designed sensitivity of Square Kilometer Array~(SKA).

Cosmological GW sources, associated with the presence of energy-density inhomogeneities verified by lattice simulations~\cite{Hindmarsh:2015qta,Cutting:2018tjt,Di:2020kbw,Hiramatsu:2013qaa,Press:1989yh}, generally also induce significant energy density perturbations in both radiation and dark matter~\cite{Liu:2023tmv,Gouttenoire:2025wxc}. After the radiation-matter equivalence, the density perturbations of DM increases with time, and the DM subhalos decouple from the expansion of the Universe when the perturbation amplitude reaches the $\mathcal{O}(1)$. 
In this $Letter$, we investigate the constraints on the popular cosmological GW sources, including first-order phase transitions, domain walls, long strings and the fragmentation of boson condensates, from the upper limits of compact subhalos, and set more strict constraints on the energy spectra of cosmological sources below the astrophysical foreground noise. 

\emph{Observing the compact DM subhalos with pulsar timing}.
The compact subhalos of cold DM emerge from enhanced power of density perturbations at small scales. 
Perturbations in the DM component can grow linearly with the scale factor after the matter-radiation equality. The small-scale enhanced perturbations reach the nonlinear evolution regime and overdense regions of DM decouple from the expansion of the Universe at an early era like the recombination ($z\sim 1000$). Generally speaking, perturbations with the amplitude $\delta  \gtrsim 10^{-3}$ at the horizon reentry can lead to abundant formation of the compact subhalos. 
Recent numerical simulations \cite{Delos:2017thv,Delos:2018ueo} revealed that the DM density in a compact subhalo originating from the spike-type power spectrum on small scales has the Moore profile, represented by 
\begin{equation}
\begin{split}
    &\rho(r) = \dfrac{\rho_s}{(r/r_s)^{3/2}(1+r/r_s)^{3/2}}\,,\\
    \rho_s \simeq 30 &(1+z_c)^3 \rho_{m,0}, \quad r_s \simeq 0.7[(1+z_c)k_s]^{-1}, 
\end{split}
\end{equation}
where $z_{c}$ is the redshift of the last collapse, $\rho_{m,0}$ is the present matter density, and $k_s$ is the peak wave number of enhanced perturbations.
The mass of the compact subhalos, $M_{\rm vir}$, is determined by the total mass enclosed within the viral radius $r_{\rm vir}$,
\begin{align}
M_{\rm vir} = 4 \pi \rho_s r_s^3 \int^{x_{\rm vir}}_0 \frac{x^{1/2}}{(1+x)^{3/2}} dx,
\end{align}
where $x\equiv r/r_s$ ($x_{\rm vir}=r_{\rm vir}/r_s$) and $r_{\rm vir}$ is defined as the radius within which the average DM density is $200$ times the mean DM density of the Universe. 
Assuming Gaussianity of the density perturbations, the fraction of DM that exist in the compact subhalos is given by 
\begin{align}
    F = \frac{M_{\rm vir}}{M_{\mathrm{i}}} \int^{0.3}_{\delta_{\rm min}} \frac{1}{\sqrt{2\pi \sigma_H^2}} \exp{\left( -\frac{\delta^2}{2\sigma_H^2} \right)}d\delta\,,
\end{align}
where the threshold $\delta_{\rm min}$ is associated with $k_s$ and $z_c$, and we determine it by following the results presented in Ref. \cite{Bringmann:2011ut}, $\sigma_H$ is the standard deviation of the smoothed DM density contrast at horizon entry, $M_{\rm i}$ is the initial mass of DM component within the overdense region, specifically the mass at the moment the perturbation with a spike-wave number $k_s$ enters the horizon, given by $M_{\rm i} \simeq \frac{4\pi}{3}k_s^{-3} \rho_{\rm DM,0}$.

Observations of compact subhalo could impose stringent constraints on primordial curvature perturbations ($P_{\mathcal{R}} \lesssim 10^{-6}$) at scales $k \lesssim 10^7$ Mpc$^{-1}$, contingent upon detecting whether dark matter predominantly resides in these structures. Such detections exploit gravitational lensing effects and DM annihilation signals. We focus on the model-independent case where cold DM only interact gravitationally with fields in the Standard Model. Since the energy scale of the GW sources simultaneously determine the halo mass and the GW peak frequency, the frequency range of the GW energy spectra maps to halo mass ranges ($10^{-12}M_\odot \lesssim M_{\mathrm{vir}} \lesssim 1M_\odot$). These measurements constrain GW energy density spectra across nHz-$\mu$Hz bands through both individual events and collective statistical analysis of second-order pulsar frequency derivatives induced by subhalo transits through Doppler modulation.

Future SKA-enhanced pulsar timing arrays could probe primordial curvature perturbations at $k \lesssim 10^6$ Mpc$^{-1}$ by resolving $\mathcal{O}(1)$ halo abundance. Using baseline parameters from Ref.~\cite{Rosado:2015epa}. ($N_p=200$, $\sigma_t=50$ ns, $\Delta t=14$-day cadence, $z_0=5$ kpc), the signal-to-noise ratio demonstrates critical temporal dependence: SNR $\propto T^{7/2}$ for transient signals and $\propto T^{3/2}$ for stochastic backgrounds, necessitating $\sim$30-year campaigns to suppress timing uncertainties. Notably, constraints remain fundamentally limited by the Gaussian tail sensitivity in subhalo abundance integrals, rendering observational improvements beyond certain thresholds ineffective for further perturbation constraints.


\emph{Constraints on gravitational waves of cosmological origin}.
For all frequency bands, cleaning the astrophysical foregrounds are important for detecting SGWBs from the cosmological origin. The foreground in the mHz band from binaries of black holes and neutron stars can be suppressed by several orders of magnitude by utilizing multiband GW observation methods. The third-generation ground based GW detectors, such as the Einstein Telescope~\cite{Punturo:2010zz} and the Cosmic Explorer~\cite{Reitze:2019iox}, provide precise measurement of the statistical properties of the GW sources in the kHz band and give reliable prediction of the foreground in the mHz band that can be deduced. However, in the nHz/$\mu$Hz bands the dominant GW source is the SMBHBs, whose number density is significantly lower than the astrophysical black holes with masses $\mathcal{O}(10)$ $M_{\odot}$. In this case, the shot noise in the foreground becomes important and motivates new methods to constrain the cosmological GW strength from the observations of compact subhalos.

Along with the expansion and the decrease of the energy scale of the Big Bang Universe, higher symmetry is spontaneously broken, resulting in phase transitions~(PT). Efficient GW production are realized from both the phase transition process and their products. First-order PTs receive extensive investigations~(see Ref.~\cite{Caprini:2015zlo,Caprini:2019egz,Weir:2017wfa,Mazumdar:2018dfl} for reviews), where the vacuum energy is released within the Hubble time and concentrates upon the bubble walls and adjacent sound waves of background plasma. The violent interactions between bubble walls and sound waves trigger large-amplitude energy density perturbations of the transverse-traceless mode and release abundant GWs. For second-order PTs, GWs can also be efficiently produced in case of a large vacuum expectation value of the scalar field. The amplitudes of perturbations are amplified exponentially from parametric resonance and extract the energy from the oscillating background condensate. This mechanism is also well-known as the hybrid inflation model. Similarly, the spectator fields during inflation roll down the effective potential as its effective mass exceeds the Hubble constant which results in a similar physical picture. The topological defects, such as domain walls and cosmic strings, generally arise after the spontaneous breaking of the discrete and $U(1)$ symmetries. The stochastic motion of these stable objects are driven by their tension with nearly the speed of light and provide long term GW sources. These phenomena have been explored as potential sources to explain the nHz SGWB discovered by NANOGrav and other PTA collaborations \cite{Ellis:2023oxs}. 

These violent physical processes after inflation also induce large curvature perturbations as another consequence, which generally peak at the scale of the typical separation of the GW sources. 
A spike-like power spectrum of the density perturbations can be induced by the randomness of the quantum tunneling process during the first-order PTs ~\cite{Liu:2022lvz}, the amplification of perturbations from parametric resonance~\cite{Cai:2018tuh,Cai:2019bmk}, or by the irregular distribution of the topological defects~\cite{Press:1989yh,Saikawa:2017hiv}.

As outlined in the preceding section, the abundance of these subhalos serves as a tool for constraining model parameters. This, in turn, allows for the refinement of SGWB predictions derived from these models. To establish a conservative bound on SGWBs, we opt for a straightforward approach by setting $F = 1 $, rather than applying the constraints on their abundance from the observations of gamma rays. Moreover, upper limits on SGWBs are derived using the assumption that all of the compact subhalos form before a manually selected redshift $z_c$. Consequently, these limits depend on $z_c$, with tighter constraints associated with smaller $z_c$. According to recent simulations \cite{Delos:2017thv,Delos:2018ueo}, the restriction to the subhalos forming at $z_c \gtrsim 1000$ does not hold, and hence we adopt a more moderate case with $z_c = 500$ for our analysis. 

The explicit expressions of the GW energy spectrum of each GW source can be found in the supplementary material. Figures~\ref{fig1} and \ref{fig3} illustrate the capability of observing the subhalos to impose constraints on the cosmological SGWBs.  The yellow violin plot represents the NANOGrav 15-yr dataset, and the yellow shaded band delineates the inferred range for the astrophysical foreground assuming that the nHz SGWB mainly originates from SMBHBs. The solid green line corresponds to the theoretically predicted spectrum of the SGWB from SMBHBs. The brown dashed line denotes the expected sensitivity curve for SKA. The red/orange solid lines depict the upper limits on the energy spectra of the cosmological SGWBs with the condition $ F\lesssim 1 $. In the following, we will discuss each cosmological GW source in detail. 

For the first-order PTs, it has been demonstrated in Ref.~\cite{Liu:2022lvz} that $\sigma_H$ is proportional to $\alpha_\ast$ across various values of $\beta/H_\ast$, and the precise relationship between $\sigma_H/\alpha_\ast$ and $\beta/H_\ast$ obtained by the numerical methods of Ref.~\cite{Liu:2022lvz}. Here, $\alpha_\ast$ and $\beta^{-1}$ represent the strength and the duration of the phase transition, respectively, see the supplemental material for details. In Fig.~\ref{fig1}, we present the upper bound on $\Omega_{\mathrm{GW}}$ from bubble collisions~(upper panel) and sound waves~(lower panel) during the FOPTs for three typical values of $\beta/H_\ast$. We present the upper bound of $\Omega_{\mathrm{GW}}$ for the FOPT energy scale ranging from $10^{-3}$GeV to $10^{2}$GeV. 
As a reference, the red dashed lines represent several cases with stronger SGWBs. However, all these cases remain weaker than the irreducible astrophysical foreground noise, which are difficult to observe solely from the GW observations. The new method imposes constraints on $\Omega_{\mathrm{GW}}$ and corresponding $\alpha$ parameters that are several orders of magnitude tighter than GW detection limits. For smaller $\beta/H_{\ast}$, the wavelength of density perturbations is more close to the horizon scale, which result in stronger constraints. For bubble collisions with $\beta/H_{\ast}=100$, the GW energy spectra lie far below the foreground noise even with $\alpha \gg 1$, making it difficult to constrain the FOPTs via the observations of GWs.

For domain walls, the amplitude of $\sigma_H$ is close to the domain-wall density contrast, $\sigma_H \simeq 2\mathcal{A}\sigma/( 3 M_{\rm pl}^2 H_{\rm ann})$, with $M_{\rm pl}$ being the reduced Planck mass. Here, $\mathcal{A} \simeq 0.8$ takes an almost constant value according to the simulation results \cite{Hiramatsu:2013qaa}, $\sigma$ denotes the tension of domain walls, and $H_{\rm ann}$ is the Hubble parameter at the domain wall annihilation. We plot the peaks of the SGWB energy spectra of the domain walls, for a range of annihilation temperature $T_{\rm ann}$ spanning from $10^{-2}$GeV to $10^{3}$GeV in the left panel of Fig.~\ref{fig3}. In the scaling solutions, domain walls naturally have curvature radius comparable to the Hubble scale. Their inhomogeneous distribution induces strong density perturbations, causing the observations of compact subhalos to impose upper limits on $\Omega_{\mathrm{GW}}$ that are typically 4-6 orders of magnitude stricter than the astrophysical foreground noise.

For the fragmentation of scalar condensates, perturbations are amplified by parametric resonance during the oscillations of the condensate, induced density perturbations in the scalar field component reach the $\mathcal{O}(1)$ level at the length scale of the mean separation of the fragments, $k_{\mathrm{res}}^{-1}$. The amplitude of perturbations is proportional to $k^{-3/2}$ because of causality so that at the Hubble horizon scale $\delta_{H}\sim \Omega_{\nabla}(\frac{k_{\mathrm{res}}}{aH_{\ast}})^{-3/2}$, where $\Omega_{\nabla}$ is the energy fraction of the gradient energy $\Omega_{\nabla}\approx\frac{1}{3}\Omega_{\mathrm{\phi}}$. In the right panel of Fig.~\ref{fig3} we depict the peak values of $\Omega_{\mathrm{GW}}$ under the condition $\Omega_{\phi}=1$,  roughly $\Omega_{\mathrm{GW}}\approx 10^{-12}$, which is about four orders of magnitude below the GW foreground.

\begin{figure*}
\centering
\includegraphics[width=1.\textwidth ]{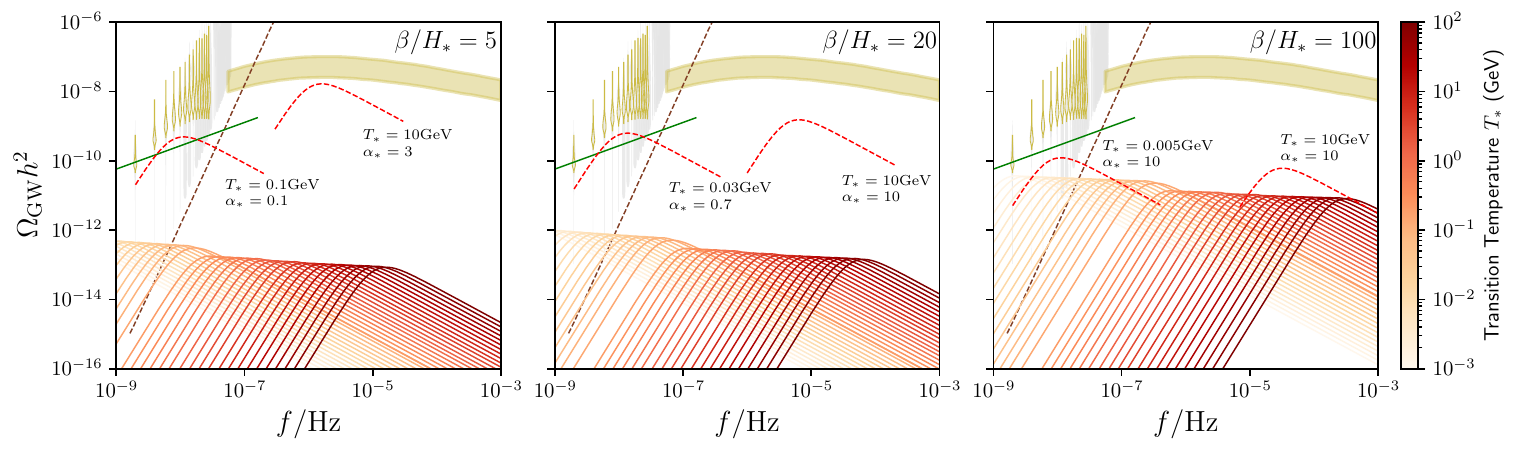}
\includegraphics[width=1.\textwidth ]{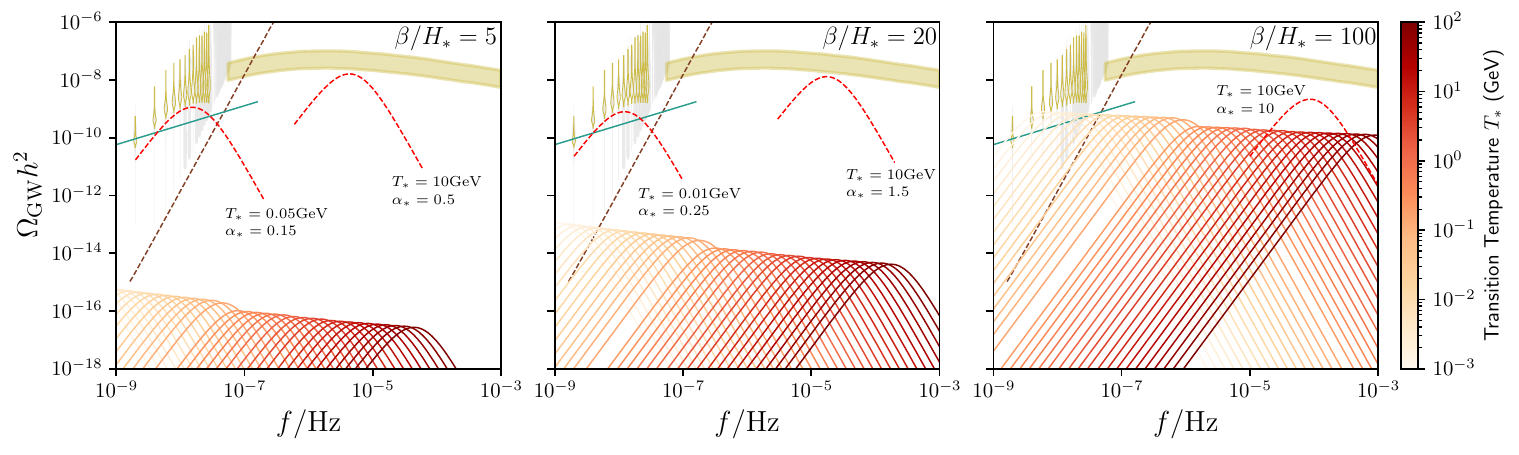}
\caption{\label{fig1} The energy spectra of SGWBs from bubble collisions~(upper panel) and sound waves~(lower panel) for $\beta/H_\ast=5$, $20$, and $100$, corresponding to $F\lesssim1$. We also depict the variations in the energy spectrum as a function of the transition temperature $T_\ast$, ranging from $10^{-3}$GeV to $10^{2}$GeV. The gray violin diagrams represent the free spectrum posteriors in the analyses from the NANOGrav 15-yr data set \cite{NANOGrav:2023hvm}. The orange-yellow violin diagrams depict the best fits to the NANOGrav 15-yr data set for SMBH binaries with environmental effects, and the orange-yellow band extends the “violins” in the PTA range to higher frequencies \cite{Ellis:2023oxs}. The green solid line shows the GW energy spectrum predicted by the VHM model of massive black hole assembly \cite{Sesana:2008mz}. The brown dashed line shows the expected sensitivity curve of the forthcoming GW experiment SKA \cite{Janssen:2014dka}.}
\end{figure*}

\begin{figure*}
\centering
\includegraphics[width=0.75\columnwidth]{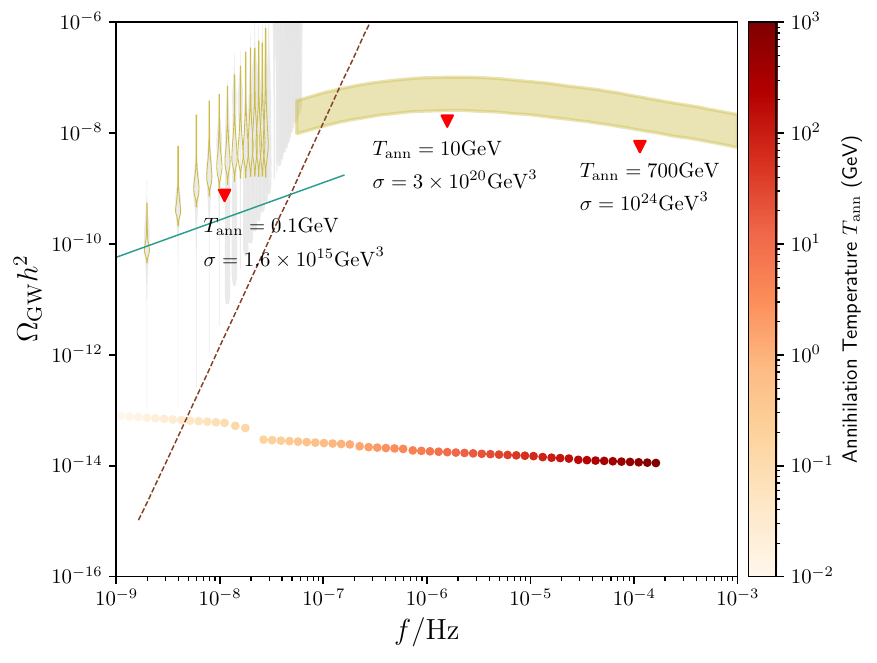}
\includegraphics[width=0.75\columnwidth]{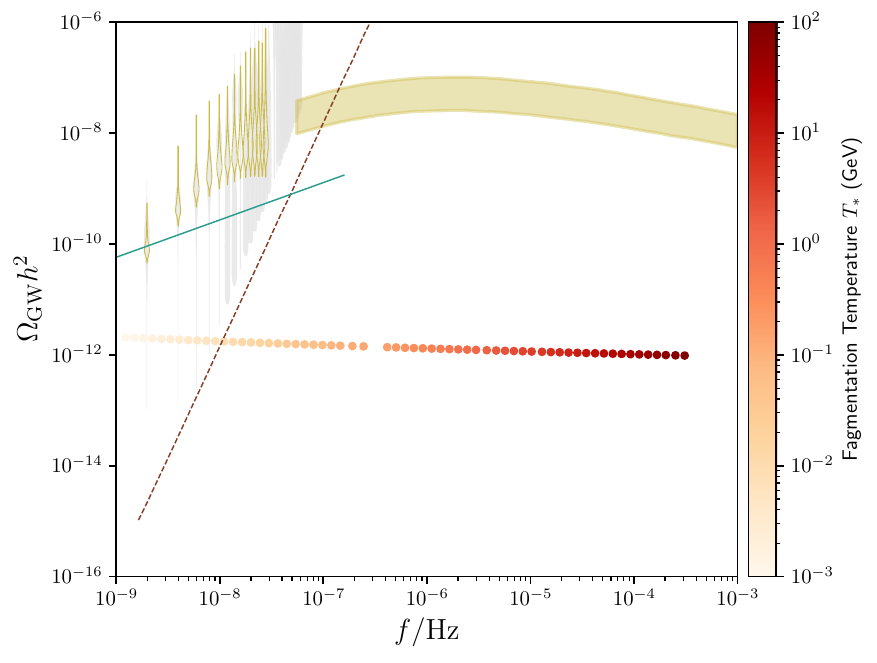}
\caption{\label{fig3} The peaks of current energy spectra of SGWBs from domain walls, corresponding to $F=1$. We also depict the variations in the peak of energy spectrum as a function of the annihilation temperature $T_{\rm ann}$, ranging from $10^{-2}$GeV to $10^{3}$GeV.}
\end{figure*}

\emph{Conclusion and discussion}. 
In this work, we propose a novel method to constrain the energy spectrum of cosmological SGWBs below the foreground of the astrophysical sources in the nHz/$\mu$Hz frequency bands through the future observations of compact subhalos of DM achieved by SKA. We investigate various prospective cosmological sources including bubble collisions, sound waves, domain walls and condensate fragmentation, and obtain the maximum $\Omega_{\mathrm{GW}}$  of each GW source under the condition that the abundance of compact subhalos is less than 1. The upper bound of $\Omega_{\mathrm{GW}}$ is generally orders of magnitude below the astrophysical foreground. Along with the development of future nHz GW observations, the astrophysical foreground noise will become more important against the measurement of cosmological SGWBs. This work provides a prospective method to constrain the GW source and corresponding new physics.

The compact subhalo formation mechanism in this work is based on the standard cold DM models.
However, this mechanism has limited applicability to fuzzy DM and warm DM models that are proposed to suppress the small-scale structures because of the large de Broglie wavelength or fast velocity of DM particles. A series of observational experiments, such as the large-scale structure, the superradiant instabilities and the direct interaction between fuzzy DM and pulsars, can help to constrain the fuzzy/warm DM models and test the validity of the cold DM model.

The upper limits of cosmological SGWBs obtained in this work can also reduce the uncertainty of the astrophysical SGWB in the nHz and $\mu$Hz bands, which yields more concrete information of the evolution and the formation rate of the SMBHBs.

Small-scale primordial curvature perturbations with large amplitude also provide potential low-frequency cosmological GW sources. These perturbations can arise from non-attractor dynamics during inflation. Curvature perturbations generate GWs through the coupling appears in the second-order expansion of the Einstein equation with the approximated expression, $\Omega^{\mathrm{pc}}_{\mathrm{GW}}\sim \Omega_{r}P_{R}^{2}$ and $\delta_{H}\sim\sqrt{P_{R}(k)}$. The constraints from compact subhalos on this kind of source is significantly tighter than other sources, typically $\Omega_{\mathrm{GW}}\lesssim10^{-17}$. 

Many non-slow-roll inflationary models predict large-amplitude primordial perturbations at small scales~(see Rev.~\cite{Ozsoy:2023ryl} for a recent review) so that scalar-induced GWs are strong enough to be observed by multiband GW detectors. The energy spectrum of scalar-induced GWs and $\delta_{H}$ depend on the specific form of the power spectrum of curvature perturbations, $P_{R}(k)$. Here, we simply apply  See Refs.~\cite{Press:1973iz,Kohri:2018awv,Espinosa:2018eve} for the explicit result. This issue has been considered before in, for example, Refs.~\cite{Delos:2023fpm,StenDelos:2022jld}, where the authors obtain constraints on $P_{R}$ from compact substructures of DM. A spiked power spectrum with fluctuations enhanced over a narrow range of scales can emerge from the inflationary phenomenology, such as the parametric resonance of curvature perturbations \cite{Cai:2018tuh,Cai:2019bmk} and the particle production during inflation \cite{Barnaby:2010ke,Cai:2021wzd}. 
 
\begin{acknowledgments}
	This work is supported in part by the National Key Research and Development Program of China Grants No. 2020YFC2201501 and No. 2021YFC2203002, in part by the National Natural Science Foundation of China Grants No. 12305057, No. 12105060, No. 12147103, No. 12235019, No. 12075297 and No. 12147103, in part by the Science Research Grants from the China Manned Space Project with NO. CMS-CSST-2021-B01, in part by the Fundamental Research Funds for the Central Universities. 
\end{acknowledgments}

\appendix
\section{The GW energy spectra}
In the supplemental material, we present the detailed expressions of the GW energy spectrum and the Hubble-scale density fluctuations from each GW source in the main text.
\subsection{Bubble collisions}
For bubble collisions and sound waves, the energy spectra of SGWBs at present from are respectively given as follows   \cite{NANOGrav:2021flc}:
\begin{align}
\Omega_{\rm GW}(f)h^2 = 3.57 \times 10^{-5} \left( \frac{g_\ast(T_\ast)}{10}\right)^{-1/3}\frac{0.48v_w^3}{1+5.3v_w^2+5v_w^4}   \nonumber \\
\times  \left( \frac{H_\ast}{\beta}\right)^2 \left( \frac{\kappa_\phi \alpha_\ast}{1+\alpha_\ast}\right)^2 \left(\frac{4}{ (f/f_{\rm p})^{-2} + 3(f/f_{\rm p})^{2/3} }\right)^{3/2}
\end{align}
with 
\begin{align}
    f_{\rm p} = 1.13 \times 10^{-10} {\rm{Hz}} \left(\frac{0.35}{1+0.07 v_w+0.69 v_w^4} \right) \nonumber \\
    \times \left( \frac{\beta}{H_\ast}\right) \left( \frac{T_\ast}{\rm MeV} \right)\left( \frac{g_\ast(T_\ast)}{10}\right)^{1/6},
\end{align}
and 
\begin{align}
\Omega_{\rm GW}(f)h^2 = &3.57 \times 10^{-5} \left( \frac{g_\ast(T_\ast)}{10}\right)^{-1/3} 0.513v_w  \left( \frac{H_\ast}{\beta}\right) \nonumber \\
&\times \left( \frac{\kappa_{\rm sw} \alpha_\ast}{1+\alpha_\ast}\right)^2  (f/f_{\rm p})^3 \left( \frac{7}{4+3(f/f_{\rm p})^2} \right)^{7/2}\nonumber \\
&\times \left[ 1 - (1+2\tau_{\rm sw}H_\ast)^{-1/2} \right]
\end{align}
with 
\begin{align}
f_{\rm p} = 1.13 \times 10^{-10} {\rm{Hz}} \left(\frac{0.536}{v_w} \right) \left( \frac{\beta}{H_\ast}\right) \left( \frac{T_\ast}{\rm MeV} \right)\left( \frac{g_\ast(T_\ast)}{10}\right)^{1/6},
\end{align} 
where $T_\ast$ represents the Universe temperature when the phase transition occurs and $g_\ast$ denotes the number of relativistic degrees of freedom for the energy density. We use $v_w$ to denote the velocity of the bubble walls, and set $v_w=1$ to obtain a conservative estimation of $\Omega_{\mathrm{GW}}$. The sound-wave lifetime $\tau_{\rm sw}$ is usually taken to be $\tau_{\rm sw} \simeq (8\pi)^{1/3}\beta^{-1}\sqrt{4(1+\alpha_\ast)/(3\kappa_{\rm sw}\alpha_\ast)}$ \cite{Weir:2017wfa}. In addition, $\kappa_\phi$ and $\kappa_{\rm sw}$ respectively denote the fraction of vacuum energy transformed into the kinetic energy of the scalar field and the bulk motion of the fluid. To obtain a conservative estimation, we conduct two separate analyses: one focusing solely on bubble collisions and another concentrating exclusively on sound waves. In the former case, we set $\kappa_{\phi}=1$ and $\kappa_{\rm sw}=0$. Conversely, in the latter case, we maintain $\kappa_{\phi}=0$ and utilize the findings from Ref. \cite{Espinosa:2010hh} to calculate $\kappa_{\rm sw} \simeq \alpha_\ast /(0.73+0.083\sqrt{\alpha_\ast}+\alpha_\ast)$.

$\quad$

\subsection{Domain walls}
The peak amplitude of GWs at the present is given by~\cite{Hiramatsu:2013qaa,Saikawa:2017hiv}
\begin{align}
\left(\Omega_{\rm GW}h^2\right)_{\rm p} = 7.2 \times 10^{-18} \tilde\epsilon_{\rm gw} \mathcal{A}^2 \left( \frac{g_{\ast}(T_{\rm ann})}{10}\right)^{-4/3} \nonumber \\
\times \left( \frac{\sigma}{1{\rm TeV}^3}\right)^{2} \left( \frac{T_{\rm ann}}{10^{-2} {\rm GeV}}\right)^{-4},
\end{align}
where $\sigma$ is the tension of domain walls, $T_{\mathrm{ann}}$ is the annihilation time of domain walls~(the $Z_2$ symmetry is explicitly broken by a bias term), and$\tilde\epsilon_{\rm gw} \simeq 0.7$ and $\mathcal{A}\approx 0.8$ are determined by the results of numerical simulations.
The peak frequency of GW energy spectrum can be estimated as 
\begin{align}
f_{\rm p}  = 1.1\times & 10^{-9} {\rm{Hz}} \left( \frac{g_{\ast}(T_{\rm ann})}{10}\right)^{1/2} \nonumber \\
&\times \left( \frac{g_{\ast s}(T_{\rm ann})}{10}\right)^{-1/3}  \left( \frac{T_{\rm ann}}{10^{-2} {\rm GeV}}\right).
\end{align}
For the infrared region~($f\ll f_\mathrm{p}$) and the ultraviolet region~($f\gg f_\mathrm{p}$), $\Omega_{\mathrm{GW}}(f)$ is proportional to $f^3$ and $f^{-1}$ respectively.

\subsection{Condensate fragmentation}
Post-inflation oscillations of the inflaton or spectator fields~(when $H < m_{\mathrm{eff}}$) result in the resonant amplification of perturbations through strong interactions.
Because of the nonlinear evolution, the GW energy spectra are obtained by lattice simulations~\cite{Kofman:1997yn,Khlebnikov:1997di,Amin:2010dc,Zhou:2013tsa,Lozanov:2019ylm,Liu:2017hua,Olle:2019kbo,Inomata:2020xad,Kitajima:2020rpm,Ramberg:2020oct,Kawasaki:2021ycf}, which are approximately written in a unified form~\cite{Lozanov:2019ylm}
    \begin{equation}
\Omega_{\mathrm{GW}}^{\mathrm{cf}}\approx 0.1\Omega_{\mathrm{r}}(\frac{k_{\mathrm{res}}}{a_{*}H_{*}})^{-2}\,,
    \end{equation}
    where $\Omega_{\mathrm{r}}$ is the present energy fraction of radiation and $k_{\mathrm{res}}$ denotes the approximate wavenumber of energy density perturbations with the maximum resonance rate. $\Omega_{\mathrm{GW}}(f)$ is proportional to $f^3$ in the infrared region and sharply decrease for $f>f_{\mathrm{p}}$ but The specific form depends on the specific models.

\bibliographystyle{apsrev4-1}
\bibliography{references}  

\begin{thebibliography}{110}%
\makeatletter
\providecommand \@ifxundefined [1]{%
 \@ifx{#1\undefined}
}%
\providecommand \@ifnum [1]{%
 \ifnum #1\expandafter \@firstoftwo
 \else \expandafter \@secondoftwo
 \fi
}%
\providecommand \@ifx [1]{%
 \ifx #1\expandafter \@firstoftwo
 \else \expandafter \@secondoftwo
 \fi
}%
\providecommand \natexlab [1]{#1}%
\providecommand \enquote  [1]{``#1''}%
\providecommand \bibnamefont  [1]{#1}%
\providecommand \bibfnamefont [1]{#1}%
\providecommand \citenamefont [1]{#1}%
\providecommand \href@noop [0]{\@secondoftwo}%
\providecommand \href [0]{\begingroup \@sanitize@url \@href}%
\providecommand \@href[1]{\@@startlink{#1}\@@href}%
\providecommand \@@href[1]{\endgroup#1\@@endlink}%
\providecommand \@sanitize@url [0]{\catcode `\\12\catcode `\$12\catcode
  `\&12\catcode `\#12\catcode `\^12\catcode `\_12\catcode `\%12\relax}%
\providecommand \@@startlink[1]{}%
\providecommand \@@endlink[0]{}%
\providecommand \url  [0]{\begingroup\@sanitize@url \@url }%
\providecommand \@url [1]{\endgroup\@href {#1}{\urlprefix }}%
\providecommand \urlprefix  [0]{URL }%
\providecommand \Eprint [0]{\href }%
\providecommand \doibase [0]{http://dx.doi.org/}%
\providecommand \selectlanguage [0]{\@gobble}%
\providecommand \bibinfo  [0]{\@secondoftwo}%
\providecommand \bibfield  [0]{\@secondoftwo}%
\providecommand \translation [1]{[#1]}%
\providecommand \BibitemOpen [0]{}%
\providecommand \bibitemStop [0]{}%
\providecommand \bibitemNoStop [0]{.\EOS\space}%
\providecommand \EOS [0]{\spacefactor3000\relax}%
\providecommand \BibitemShut  [1]{\csname bibitem#1\endcsname}%
\let\auto@bib@innerbib\@empty
\bibitem [{\citenamefont {Caprini}(2015)}]{Caprini:2015tfa}%
  \BibitemOpen
  \bibfield  {author} {\bibinfo {author} {\bibfnamefont {C.}~\bibnamefont
  {Caprini}},\ }\href {\doibase 10.1088/1742-6596/610/1/012004} {\bibfield
  {journal} {\bibinfo  {journal} {J. Phys. Conf. Ser.}\ }\textbf {\bibinfo
  {volume} {610}},\ \bibinfo {pages} {012004} (\bibinfo {year} {2015})},\
  \Eprint {http://arxiv.org/abs/1501.01174} {arXiv:1501.01174 [gr-qc]}
  \BibitemShut {NoStop}%
\bibitem [{\citenamefont {Kuroyanagi}\ \emph {et~al.}(2018)\citenamefont
  {Kuroyanagi}, \citenamefont {Chiba},\ and\ \citenamefont
  {Takahashi}}]{Kuroyanagi:2018csn}%
  \BibitemOpen
  \bibfield  {author} {\bibinfo {author} {\bibfnamefont {S.}~\bibnamefont
  {Kuroyanagi}}, \bibinfo {author} {\bibfnamefont {T.}~\bibnamefont {Chiba}}, \
  and\ \bibinfo {author} {\bibfnamefont {T.}~\bibnamefont {Takahashi}},\ }\href
  {\doibase 10.1088/1475-7516/2018/11/038} {\bibfield  {journal} {\bibinfo
  {journal} {JCAP}\ }\textbf {\bibinfo {volume} {11}},\ \bibinfo {pages} {038}
  (\bibinfo {year} {2018})},\ \Eprint {http://arxiv.org/abs/1807.00786}
  {arXiv:1807.00786 [astro-ph.CO]} \BibitemShut {NoStop}%
\bibitem [{\citenamefont {Caprini}\ and\ \citenamefont
  {Figueroa}(2018)}]{Caprini:2018mtu}%
  \BibitemOpen
  \bibfield  {author} {\bibinfo {author} {\bibfnamefont {C.}~\bibnamefont
  {Caprini}}\ and\ \bibinfo {author} {\bibfnamefont {D.~G.}\ \bibnamefont
  {Figueroa}},\ }\href {\doibase 10.1088/1361-6382/aac608} {\bibfield
  {journal} {\bibinfo  {journal} {Class. Quant. Grav.}\ }\textbf {\bibinfo
  {volume} {35}},\ \bibinfo {pages} {163001} (\bibinfo {year} {2018})},\
  \Eprint {http://arxiv.org/abs/1801.04268} {arXiv:1801.04268 [astro-ph.CO]}
  \BibitemShut {NoStop}%
\bibitem [{\citenamefont {Christensen}(2019)}]{Christensen:2018iqi}%
  \BibitemOpen
  \bibfield  {author} {\bibinfo {author} {\bibfnamefont {N.}~\bibnamefont
  {Christensen}},\ }\href {\doibase 10.1088/1361-6633/aae6b5} {\bibfield
  {journal} {\bibinfo  {journal} {Rept. Prog. Phys.}\ }\textbf {\bibinfo
  {volume} {82}},\ \bibinfo {pages} {016903} (\bibinfo {year} {2019})},\
  \Eprint {http://arxiv.org/abs/1811.08797} {arXiv:1811.08797 [gr-qc]}
  \BibitemShut {NoStop}%
\bibitem [{\citenamefont {Renzini}\ \emph {et~al.}(2022)\citenamefont
  {Renzini}, \citenamefont {Goncharov}, \citenamefont {Jenkins},\ and\
  \citenamefont {Meyers}}]{Renzini:2022alw}%
  \BibitemOpen
  \bibfield  {author} {\bibinfo {author} {\bibfnamefont {A.~I.}\ \bibnamefont
  {Renzini}}, \bibinfo {author} {\bibfnamefont {B.}~\bibnamefont {Goncharov}},
  \bibinfo {author} {\bibfnamefont {A.~C.}\ \bibnamefont {Jenkins}}, \ and\
  \bibinfo {author} {\bibfnamefont {P.~M.}\ \bibnamefont {Meyers}},\ }\href
  {\doibase 10.3390/galaxies10010034} {\bibfield  {journal} {\bibinfo
  {journal} {Galaxies}\ }\textbf {\bibinfo {volume} {10}},\ \bibinfo {pages}
  {34} (\bibinfo {year} {2022})},\ \Eprint {http://arxiv.org/abs/2202.00178}
  {arXiv:2202.00178 [gr-qc]} \BibitemShut {NoStop}%
\bibitem [{\citenamefont {Agazie}\ \emph
  {et~al.}(2023{\natexlab{a}})\citenamefont {Agazie} \emph
  {et~al.}}]{NANOGrav:2023gor}%
  \BibitemOpen
  \bibfield  {author} {\bibinfo {author} {\bibfnamefont {G.}~\bibnamefont
  {Agazie}} \emph {et~al.} (\bibinfo {collaboration} {NANOGrav}),\ }\href
  {\doibase 10.3847/2041-8213/acdac6} {\bibfield  {journal} {\bibinfo
  {journal} {Astrophys. J. Lett.}\ }\textbf {\bibinfo {volume} {951}},\
  \bibinfo {pages} {L8} (\bibinfo {year} {2023}{\natexlab{a}})},\ \Eprint
  {http://arxiv.org/abs/2306.16213} {arXiv:2306.16213 [astro-ph.HE]}
  \BibitemShut {NoStop}%
\bibitem [{\citenamefont {Agazie}\ \emph
  {et~al.}(2023{\natexlab{b}})\citenamefont {Agazie} \emph
  {et~al.}}]{NANOGrav:2023hde}%
  \BibitemOpen
  \bibfield  {author} {\bibinfo {author} {\bibfnamefont {G.}~\bibnamefont
  {Agazie}} \emph {et~al.} (\bibinfo {collaboration} {NANOGrav}),\ }\href
  {\doibase 10.3847/2041-8213/acda9a} {\bibfield  {journal} {\bibinfo
  {journal} {Astrophys. J. Lett.}\ }\textbf {\bibinfo {volume} {951}},\
  \bibinfo {pages} {L9} (\bibinfo {year} {2023}{\natexlab{b}})},\ \Eprint
  {http://arxiv.org/abs/2306.16217} {arXiv:2306.16217 [astro-ph.HE]}
  \BibitemShut {NoStop}%
\bibitem [{\citenamefont {Antoniadis}\ \emph
  {et~al.}(2023{\natexlab{a}})\citenamefont {Antoniadis} \emph
  {et~al.}}]{EPTA:2023fyk}%
  \BibitemOpen
  \bibfield  {author} {\bibinfo {author} {\bibfnamefont {J.}~\bibnamefont
  {Antoniadis}} \emph {et~al.} (\bibinfo {collaboration} {EPTA, InPTA:}),\
  }\href {\doibase 10.1051/0004-6361/202346844} {\bibfield  {journal} {\bibinfo
   {journal} {Astron. Astrophys.}\ }\textbf {\bibinfo {volume} {678}},\
  \bibinfo {pages} {A50} (\bibinfo {year} {2023}{\natexlab{a}})},\ \Eprint
  {http://arxiv.org/abs/2306.16214} {arXiv:2306.16214 [astro-ph.HE]}
  \BibitemShut {NoStop}%
\bibitem [{\citenamefont {Antoniadis}\ \emph
  {et~al.}(2023{\natexlab{b}})\citenamefont {Antoniadis} \emph
  {et~al.}}]{EPTA:2023sfo}%
  \BibitemOpen
  \bibfield  {author} {\bibinfo {author} {\bibfnamefont {J.}~\bibnamefont
  {Antoniadis}} \emph {et~al.} (\bibinfo {collaboration} {EPTA}),\ }\href
  {\doibase 10.1051/0004-6361/202346841} {\bibfield  {journal} {\bibinfo
  {journal} {Astron. Astrophys.}\ }\textbf {\bibinfo {volume} {678}},\ \bibinfo
  {pages} {A48} (\bibinfo {year} {2023}{\natexlab{b}})},\ \Eprint
  {http://arxiv.org/abs/2306.16224} {arXiv:2306.16224 [astro-ph.HE]}
  \BibitemShut {NoStop}%
\bibitem [{\citenamefont {Antoniadis}\ \emph
  {et~al.}(2023{\natexlab{c}})\citenamefont {Antoniadis} \emph
  {et~al.}}]{EPTA:2023xxk}%
  \BibitemOpen
  \bibfield  {author} {\bibinfo {author} {\bibfnamefont {J.}~\bibnamefont
  {Antoniadis}} \emph {et~al.} (\bibinfo {collaboration} {EPTA}),\ }\href@noop
  {} {\  (\bibinfo {year} {2023}{\natexlab{c}})},\ \Eprint
  {http://arxiv.org/abs/2306.16227} {arXiv:2306.16227 [astro-ph.CO]}
  \BibitemShut {NoStop}%
\bibitem [{\citenamefont {Reardon}\ \emph {et~al.}(2023)\citenamefont {Reardon}
  \emph {et~al.}}]{Reardon:2023gzh}%
  \BibitemOpen
  \bibfield  {author} {\bibinfo {author} {\bibfnamefont {D.~J.}\ \bibnamefont
  {Reardon}} \emph {et~al.},\ }\href {\doibase 10.3847/2041-8213/acdd02}
  {\bibfield  {journal} {\bibinfo  {journal} {Astrophys. J. Lett.}\ }\textbf
  {\bibinfo {volume} {951}},\ \bibinfo {pages} {L6} (\bibinfo {year} {2023})},\
  \Eprint {http://arxiv.org/abs/2306.16215} {arXiv:2306.16215 [astro-ph.HE]}
  \BibitemShut {NoStop}%
\bibitem [{\citenamefont {Xu}\ \emph {et~al.}(2023)\citenamefont {Xu} \emph
  {et~al.}}]{Xu:2023wog}%
  \BibitemOpen
  \bibfield  {author} {\bibinfo {author} {\bibfnamefont {H.}~\bibnamefont {Xu}}
  \emph {et~al.},\ }\href {\doibase 10.1088/1674-4527/acdfa5} {\bibfield
  {journal} {\bibinfo  {journal} {Res. Astron. Astrophys.}\ }\textbf {\bibinfo
  {volume} {23}},\ \bibinfo {pages} {075024} (\bibinfo {year} {2023})},\
  \Eprint {http://arxiv.org/abs/2306.16216} {arXiv:2306.16216 [astro-ph.HE]}
  \BibitemShut {NoStop}%
\bibitem [{\citenamefont {Agazie}\ \emph
  {et~al.}(2023{\natexlab{c}})\citenamefont {Agazie} \emph
  {et~al.}}]{NANOGrav:2023hfp}%
  \BibitemOpen
  \bibfield  {author} {\bibinfo {author} {\bibfnamefont {G.}~\bibnamefont
  {Agazie}} \emph {et~al.} (\bibinfo {collaboration} {NANOGrav}),\ }\href
  {\doibase 10.3847/2041-8213/ace18b} {\bibfield  {journal} {\bibinfo
  {journal} {Astrophys. J. Lett.}\ }\textbf {\bibinfo {volume} {952}},\
  \bibinfo {pages} {L37} (\bibinfo {year} {2023}{\natexlab{c}})},\ \Eprint
  {http://arxiv.org/abs/2306.16220} {arXiv:2306.16220 [astro-ph.HE]}
  \BibitemShut {NoStop}%
\bibitem [{\citenamefont {Ellis}\ \emph
  {et~al.}(2024{\natexlab{a}})\citenamefont {Ellis}, \citenamefont {Fairbairn},
  \citenamefont {H\"utsi}, \citenamefont {Raidal}, \citenamefont {Urrutia},
  \citenamefont {Vaskonen},\ and\ \citenamefont {Veerm\"ae}}]{Ellis:2023dgf}%
  \BibitemOpen
  \bibfield  {author} {\bibinfo {author} {\bibfnamefont {J.}~\bibnamefont
  {Ellis}}, \bibinfo {author} {\bibfnamefont {M.}~\bibnamefont {Fairbairn}},
  \bibinfo {author} {\bibfnamefont {G.}~\bibnamefont {H\"utsi}}, \bibinfo
  {author} {\bibfnamefont {J.}~\bibnamefont {Raidal}}, \bibinfo {author}
  {\bibfnamefont {J.}~\bibnamefont {Urrutia}}, \bibinfo {author} {\bibfnamefont
  {V.}~\bibnamefont {Vaskonen}}, \ and\ \bibinfo {author} {\bibfnamefont
  {H.}~\bibnamefont {Veerm\"ae}},\ }\href {\doibase
  10.1103/PhysRevD.109.L021302} {\bibfield  {journal} {\bibinfo  {journal}
  {Phys. Rev. D}\ }\textbf {\bibinfo {volume} {109}},\ \bibinfo {pages}
  {L021302} (\bibinfo {year} {2024}{\natexlab{a}})},\ \Eprint
  {http://arxiv.org/abs/2306.17021} {arXiv:2306.17021 [astro-ph.CO]}
  \BibitemShut {NoStop}%
\bibitem [{\citenamefont {Shen}\ \emph {et~al.}(2023)\citenamefont {Shen},
  \citenamefont {Yuan}, \citenamefont {Wang},\ and\ \citenamefont
  {Wang}}]{Shen:2023pan}%
  \BibitemOpen
  \bibfield  {author} {\bibinfo {author} {\bibfnamefont {Z.-Q.}\ \bibnamefont
  {Shen}}, \bibinfo {author} {\bibfnamefont {G.-W.}\ \bibnamefont {Yuan}},
  \bibinfo {author} {\bibfnamefont {Y.-Y.}\ \bibnamefont {Wang}}, \ and\
  \bibinfo {author} {\bibfnamefont {Y.-Z.}\ \bibnamefont {Wang}},\ }\href@noop
  {} {\  (\bibinfo {year} {2023})},\ \Eprint {http://arxiv.org/abs/2306.17143}
  {arXiv:2306.17143 [astro-ph.HE]} \BibitemShut {NoStop}%
\bibitem [{\citenamefont {Ghoshal}\ and\ \citenamefont
  {Strumia}(2024)}]{Ghoshal:2023fhh}%
  \BibitemOpen
  \bibfield  {author} {\bibinfo {author} {\bibfnamefont {A.}~\bibnamefont
  {Ghoshal}}\ and\ \bibinfo {author} {\bibfnamefont {A.}~\bibnamefont
  {Strumia}},\ }\href {\doibase 10.1088/1475-7516/2024/02/054} {\bibfield
  {journal} {\bibinfo  {journal} {JCAP}\ }\textbf {\bibinfo {volume} {02}},\
  \bibinfo {pages} {054} (\bibinfo {year} {2024})},\ \Eprint
  {http://arxiv.org/abs/2306.17158} {arXiv:2306.17158 [astro-ph.CO]}
  \BibitemShut {NoStop}%
\bibitem [{\citenamefont {Bi}\ \emph {et~al.}(2023)\citenamefont {Bi},
  \citenamefont {Wu}, \citenamefont {Chen},\ and\ \citenamefont
  {Huang}}]{Bi:2023tib}%
  \BibitemOpen
  \bibfield  {author} {\bibinfo {author} {\bibfnamefont {Y.-C.}\ \bibnamefont
  {Bi}}, \bibinfo {author} {\bibfnamefont {Y.-M.}\ \bibnamefont {Wu}}, \bibinfo
  {author} {\bibfnamefont {Z.-C.}\ \bibnamefont {Chen}}, \ and\ \bibinfo
  {author} {\bibfnamefont {Q.-G.}\ \bibnamefont {Huang}},\ }\href {\doibase
  10.1007/s11433-023-2252-4} {\bibfield  {journal} {\bibinfo  {journal} {Sci.
  China Phys. Mech. Astron.}\ }\textbf {\bibinfo {volume} {66}},\ \bibinfo
  {pages} {120402} (\bibinfo {year} {2023})},\ \Eprint
  {http://arxiv.org/abs/2307.00722} {arXiv:2307.00722 [astro-ph.CO]}
  \BibitemShut {NoStop}%
\bibitem [{\citenamefont {Zhang}\ \emph
  {et~al.}(2023{\natexlab{a}})\citenamefont {Zhang}, \citenamefont {Dai},
  \citenamefont {Gao}, \citenamefont {Gong}, \citenamefont {Jiang},\ and\
  \citenamefont {Lu}}]{Zhang:2023lzt}%
  \BibitemOpen
  \bibfield  {author} {\bibinfo {author} {\bibfnamefont {C.}~\bibnamefont
  {Zhang}}, \bibinfo {author} {\bibfnamefont {N.}~\bibnamefont {Dai}}, \bibinfo
  {author} {\bibfnamefont {Q.}~\bibnamefont {Gao}}, \bibinfo {author}
  {\bibfnamefont {Y.}~\bibnamefont {Gong}}, \bibinfo {author} {\bibfnamefont
  {T.}~\bibnamefont {Jiang}}, \ and\ \bibinfo {author} {\bibfnamefont
  {X.}~\bibnamefont {Lu}},\ }\href {\doibase 10.1103/PhysRevD.108.104069}
  {\bibfield  {journal} {\bibinfo  {journal} {Phys. Rev. D}\ }\textbf {\bibinfo
  {volume} {108}},\ \bibinfo {pages} {104069} (\bibinfo {year}
  {2023}{\natexlab{a}})},\ \Eprint {http://arxiv.org/abs/2307.01093}
  {arXiv:2307.01093 [gr-qc]} \BibitemShut {NoStop}%
\bibitem [{\citenamefont {Fujikura}\ \emph {et~al.}(2023)\citenamefont
  {Fujikura}, \citenamefont {Girmohanta}, \citenamefont {Nakai},\ and\
  \citenamefont {Suzuki}}]{Fujikura:2023lkn}%
  \BibitemOpen
  \bibfield  {author} {\bibinfo {author} {\bibfnamefont {K.}~\bibnamefont
  {Fujikura}}, \bibinfo {author} {\bibfnamefont {S.}~\bibnamefont
  {Girmohanta}}, \bibinfo {author} {\bibfnamefont {Y.}~\bibnamefont {Nakai}}, \
  and\ \bibinfo {author} {\bibfnamefont {M.}~\bibnamefont {Suzuki}},\ }\href
  {\doibase 10.1016/j.physletb.2023.138203} {\bibfield  {journal} {\bibinfo
  {journal} {Phys. Lett. B}\ }\textbf {\bibinfo {volume} {846}},\ \bibinfo
  {pages} {138203} (\bibinfo {year} {2023})},\ \Eprint
  {http://arxiv.org/abs/2306.17086} {arXiv:2306.17086 [hep-ph]} \BibitemShut
  {NoStop}%
\bibitem [{\citenamefont {Addazi}\ \emph {et~al.}(2024)\citenamefont {Addazi},
  \citenamefont {Cai}, \citenamefont {Marciano},\ and\ \citenamefont
  {Visinelli}}]{Addazi:2023jvg}%
  \BibitemOpen
  \bibfield  {author} {\bibinfo {author} {\bibfnamefont {A.}~\bibnamefont
  {Addazi}}, \bibinfo {author} {\bibfnamefont {Y.-F.}\ \bibnamefont {Cai}},
  \bibinfo {author} {\bibfnamefont {A.}~\bibnamefont {Marciano}}, \ and\
  \bibinfo {author} {\bibfnamefont {L.}~\bibnamefont {Visinelli}},\ }\href
  {\doibase 10.1103/PhysRevD.109.015028} {\bibfield  {journal} {\bibinfo
  {journal} {Phys. Rev. D}\ }\textbf {\bibinfo {volume} {109}},\ \bibinfo
  {pages} {015028} (\bibinfo {year} {2024})},\ \Eprint
  {http://arxiv.org/abs/2306.17205} {arXiv:2306.17205 [astro-ph.CO]}
  \BibitemShut {NoStop}%
\bibitem [{\citenamefont {Megias}\ \emph {et~al.}(2023)\citenamefont {Megias},
  \citenamefont {Nardini},\ and\ \citenamefont {Quiros}}]{Megias:2023kiy}%
  \BibitemOpen
  \bibfield  {author} {\bibinfo {author} {\bibfnamefont {E.}~\bibnamefont
  {Megias}}, \bibinfo {author} {\bibfnamefont {G.}~\bibnamefont {Nardini}}, \
  and\ \bibinfo {author} {\bibfnamefont {M.}~\bibnamefont {Quiros}},\ }\href
  {\doibase 10.1103/PhysRevD.108.095017} {\bibfield  {journal} {\bibinfo
  {journal} {Phys. Rev. D}\ }\textbf {\bibinfo {volume} {108}},\ \bibinfo
  {pages} {095017} (\bibinfo {year} {2023})},\ \Eprint
  {http://arxiv.org/abs/2306.17071} {arXiv:2306.17071 [hep-ph]} \BibitemShut
  {NoStop}%
\bibitem [{\citenamefont {Xiao}\ \emph {et~al.}(2023)\citenamefont {Xiao},
  \citenamefont {Yang},\ and\ \citenamefont {Zhang}}]{Xiao:2023dbb}%
  \BibitemOpen
  \bibfield  {author} {\bibinfo {author} {\bibfnamefont {Y.}~\bibnamefont
  {Xiao}}, \bibinfo {author} {\bibfnamefont {J.~M.}\ \bibnamefont {Yang}}, \
  and\ \bibinfo {author} {\bibfnamefont {Y.}~\bibnamefont {Zhang}},\ }\href
  {\doibase 10.1016/j.scib.2023.11.025} {\bibfield  {journal} {\bibinfo
  {journal} {Sci. Bull.}\ }\textbf {\bibinfo {volume} {68}},\ \bibinfo {pages}
  {3158} (\bibinfo {year} {2023})},\ \Eprint {http://arxiv.org/abs/2307.01072}
  {arXiv:2307.01072 [hep-ph]} \BibitemShut {NoStop}%
\bibitem [{\citenamefont {Li}\ and\ \citenamefont {Xie}(2023)}]{Li:2023bxy}%
  \BibitemOpen
  \bibfield  {author} {\bibinfo {author} {\bibfnamefont {S.-P.}\ \bibnamefont
  {Li}}\ and\ \bibinfo {author} {\bibfnamefont {K.-P.}\ \bibnamefont {Xie}},\
  }\href {\doibase 10.1103/PhysRevD.108.055018} {\bibfield  {journal} {\bibinfo
   {journal} {Phys. Rev. D}\ }\textbf {\bibinfo {volume} {108}},\ \bibinfo
  {pages} {055018} (\bibinfo {year} {2023})},\ \Eprint
  {http://arxiv.org/abs/2307.01086} {arXiv:2307.01086 [hep-ph]} \BibitemShut
  {NoStop}%
\bibitem [{\citenamefont {Ahmadvand}\ \emph {et~al.}(2023)\citenamefont
  {Ahmadvand}, \citenamefont {Bian},\ and\ \citenamefont
  {Shakeri}}]{Ahmadvand:2023lpp}%
  \BibitemOpen
  \bibfield  {author} {\bibinfo {author} {\bibfnamefont {M.}~\bibnamefont
  {Ahmadvand}}, \bibinfo {author} {\bibfnamefont {L.}~\bibnamefont {Bian}}, \
  and\ \bibinfo {author} {\bibfnamefont {S.}~\bibnamefont {Shakeri}},\ }\href
  {\doibase 10.1103/PhysRevD.108.115020} {\bibfield  {journal} {\bibinfo
  {journal} {Phys. Rev. D}\ }\textbf {\bibinfo {volume} {108}},\ \bibinfo
  {pages} {115020} (\bibinfo {year} {2023})},\ \Eprint
  {http://arxiv.org/abs/2307.12385} {arXiv:2307.12385 [hep-ph]} \BibitemShut
  {NoStop}%
\bibitem [{\citenamefont {Gouttenoire}(2023)}]{Gouttenoire:2023bqy}%
  \BibitemOpen
  \bibfield  {author} {\bibinfo {author} {\bibfnamefont {Y.}~\bibnamefont
  {Gouttenoire}},\ }\href {\doibase 10.1103/PhysRevLett.131.171404} {\bibfield
  {journal} {\bibinfo  {journal} {Phys. Rev. Lett.}\ }\textbf {\bibinfo
  {volume} {131}},\ \bibinfo {pages} {171404} (\bibinfo {year} {2023})},\
  \Eprint {http://arxiv.org/abs/2307.04239} {arXiv:2307.04239 [hep-ph]}
  \BibitemShut {NoStop}%
\bibitem [{\citenamefont {Blasi}\ \emph {et~al.}(2023)\citenamefont {Blasi},
  \citenamefont {Mariotti}, \citenamefont {Rase},\ and\ \citenamefont
  {Sevrin}}]{Blasi:2023sej}%
  \BibitemOpen
  \bibfield  {author} {\bibinfo {author} {\bibfnamefont {S.}~\bibnamefont
  {Blasi}}, \bibinfo {author} {\bibfnamefont {A.}~\bibnamefont {Mariotti}},
  \bibinfo {author} {\bibfnamefont {A.}~\bibnamefont {Rase}}, \ and\ \bibinfo
  {author} {\bibfnamefont {A.}~\bibnamefont {Sevrin}},\ }\href {\doibase
  10.1007/JHEP11(2023)169} {\bibfield  {journal} {\bibinfo  {journal} {JHEP}\
  }\textbf {\bibinfo {volume} {11}},\ \bibinfo {pages} {169} (\bibinfo {year}
  {2023})},\ \Eprint {http://arxiv.org/abs/2306.17830} {arXiv:2306.17830
  [hep-ph]} \BibitemShut {NoStop}%
\bibitem [{\citenamefont {Barman}\ \emph {et~al.}(2023)\citenamefont {Barman},
  \citenamefont {Borah}, \citenamefont {Jyoti~Das},\ and\ \citenamefont
  {Saha}}]{Barman:2023fad}%
  \BibitemOpen
  \bibfield  {author} {\bibinfo {author} {\bibfnamefont {B.}~\bibnamefont
  {Barman}}, \bibinfo {author} {\bibfnamefont {D.}~\bibnamefont {Borah}},
  \bibinfo {author} {\bibfnamefont {S.}~\bibnamefont {Jyoti~Das}}, \ and\
  \bibinfo {author} {\bibfnamefont {I.}~\bibnamefont {Saha}},\ }\href {\doibase
  10.1088/1475-7516/2023/10/053} {\bibfield  {journal} {\bibinfo  {journal}
  {JCAP}\ }\textbf {\bibinfo {volume} {10}},\ \bibinfo {pages} {053} (\bibinfo
  {year} {2023})},\ \Eprint {http://arxiv.org/abs/2307.00656} {arXiv:2307.00656
  [hep-ph]} \BibitemShut {NoStop}%
\bibitem [{\citenamefont {Babichev}\ \emph {et~al.}(2023)\citenamefont
  {Babichev}, \citenamefont {Gorbunov}, \citenamefont {Ramazanov},
  \citenamefont {Samanta},\ and\ \citenamefont {Vikman}}]{Babichev:2023pbf}%
  \BibitemOpen
  \bibfield  {author} {\bibinfo {author} {\bibfnamefont {E.}~\bibnamefont
  {Babichev}}, \bibinfo {author} {\bibfnamefont {D.}~\bibnamefont {Gorbunov}},
  \bibinfo {author} {\bibfnamefont {S.}~\bibnamefont {Ramazanov}}, \bibinfo
  {author} {\bibfnamefont {R.}~\bibnamefont {Samanta}}, \ and\ \bibinfo
  {author} {\bibfnamefont {A.}~\bibnamefont {Vikman}},\ }\href {\doibase
  10.1103/PhysRevD.108.123529} {\bibfield  {journal} {\bibinfo  {journal}
  {Phys. Rev. D}\ }\textbf {\bibinfo {volume} {108}},\ \bibinfo {pages}
  {123529} (\bibinfo {year} {2023})},\ \Eprint
  {http://arxiv.org/abs/2307.04582} {arXiv:2307.04582 [hep-ph]} \BibitemShut
  {NoStop}%
\bibitem [{\citenamefont {Gelmini}\ and\ \citenamefont
  {Hyman}(2024)}]{Gelmini:2023kvo}%
  \BibitemOpen
  \bibfield  {author} {\bibinfo {author} {\bibfnamefont {G.~B.}\ \bibnamefont
  {Gelmini}}\ and\ \bibinfo {author} {\bibfnamefont {J.}~\bibnamefont
  {Hyman}},\ }\href {\doibase 10.1016/j.physletb.2023.138356} {\bibfield
  {journal} {\bibinfo  {journal} {Phys. Lett. B}\ }\textbf {\bibinfo {volume}
  {848}},\ \bibinfo {pages} {138356} (\bibinfo {year} {2024})},\ \Eprint
  {http://arxiv.org/abs/2307.07665} {arXiv:2307.07665 [hep-ph]} \BibitemShut
  {NoStop}%
\bibitem [{\citenamefont {Zhang}\ \emph
  {et~al.}(2023{\natexlab{b}})\citenamefont {Zhang}, \citenamefont {Cai},
  \citenamefont {Su}, \citenamefont {Wang}, \citenamefont {Yu},\ and\
  \citenamefont {Zhang}}]{Zhang:2023nrs}%
  \BibitemOpen
  \bibfield  {author} {\bibinfo {author} {\bibfnamefont {Z.}~\bibnamefont
  {Zhang}}, \bibinfo {author} {\bibfnamefont {C.}~\bibnamefont {Cai}}, \bibinfo
  {author} {\bibfnamefont {Y.-H.}\ \bibnamefont {Su}}, \bibinfo {author}
  {\bibfnamefont {S.}~\bibnamefont {Wang}}, \bibinfo {author} {\bibfnamefont
  {Z.-H.}\ \bibnamefont {Yu}}, \ and\ \bibinfo {author} {\bibfnamefont {H.-H.}\
  \bibnamefont {Zhang}},\ }\href {\doibase 10.1103/PhysRevD.108.095037}
  {\bibfield  {journal} {\bibinfo  {journal} {Phys. Rev. D}\ }\textbf {\bibinfo
  {volume} {108}},\ \bibinfo {pages} {095037} (\bibinfo {year}
  {2023}{\natexlab{b}})},\ \Eprint {http://arxiv.org/abs/2307.11495}
  {arXiv:2307.11495 [hep-ph]} \BibitemShut {NoStop}%
\bibitem [{\citenamefont {Ellis}\ \emph {et~al.}(2023)\citenamefont {Ellis},
  \citenamefont {Lewicki}, \citenamefont {Lin},\ and\ \citenamefont
  {Vaskonen}}]{Ellis:2023tsl}%
  \BibitemOpen
  \bibfield  {author} {\bibinfo {author} {\bibfnamefont {J.}~\bibnamefont
  {Ellis}}, \bibinfo {author} {\bibfnamefont {M.}~\bibnamefont {Lewicki}},
  \bibinfo {author} {\bibfnamefont {C.}~\bibnamefont {Lin}}, \ and\ \bibinfo
  {author} {\bibfnamefont {V.}~\bibnamefont {Vaskonen}},\ }\href {\doibase
  10.1103/PhysRevD.108.103511} {\bibfield  {journal} {\bibinfo  {journal}
  {Phys. Rev. D}\ }\textbf {\bibinfo {volume} {108}},\ \bibinfo {pages}
  {103511} (\bibinfo {year} {2023})},\ \Eprint
  {http://arxiv.org/abs/2306.17147} {arXiv:2306.17147 [astro-ph.CO]}
  \BibitemShut {NoStop}%
\bibitem [{\citenamefont {Kitajima}\ and\ \citenamefont
  {Nakayama}(2023)}]{Kitajima:2023vre}%
  \BibitemOpen
  \bibfield  {author} {\bibinfo {author} {\bibfnamefont {N.}~\bibnamefont
  {Kitajima}}\ and\ \bibinfo {author} {\bibfnamefont {K.}~\bibnamefont
  {Nakayama}},\ }\href {\doibase 10.1016/j.physletb.2023.138213} {\bibfield
  {journal} {\bibinfo  {journal} {Phys. Lett. B}\ }\textbf {\bibinfo {volume}
  {846}},\ \bibinfo {pages} {138213} (\bibinfo {year} {2023})},\ \Eprint
  {http://arxiv.org/abs/2306.17390} {arXiv:2306.17390 [hep-ph]} \BibitemShut
  {NoStop}%
\bibitem [{\citenamefont {Wang}\ \emph {et~al.}(2023)\citenamefont {Wang},
  \citenamefont {Lei}, \citenamefont {Jiao}, \citenamefont {Feng},\ and\
  \citenamefont {Fan}}]{Wang:2023len}%
  \BibitemOpen
  \bibfield  {author} {\bibinfo {author} {\bibfnamefont {Z.}~\bibnamefont
  {Wang}}, \bibinfo {author} {\bibfnamefont {L.}~\bibnamefont {Lei}}, \bibinfo
  {author} {\bibfnamefont {H.}~\bibnamefont {Jiao}}, \bibinfo {author}
  {\bibfnamefont {L.}~\bibnamefont {Feng}}, \ and\ \bibinfo {author}
  {\bibfnamefont {Y.-Z.}\ \bibnamefont {Fan}},\ }\href {\doibase
  10.1007/s11433-023-2262-0} {\bibfield  {journal} {\bibinfo  {journal} {Sci.
  China Phys. Mech. Astron.}\ }\textbf {\bibinfo {volume} {66}},\ \bibinfo
  {pages} {120403} (\bibinfo {year} {2023})},\ \Eprint
  {http://arxiv.org/abs/2306.17150} {arXiv:2306.17150 [astro-ph.HE]}
  \BibitemShut {NoStop}%
\bibitem [{\citenamefont {Lazarides}\ \emph {et~al.}(2023)\citenamefont
  {Lazarides}, \citenamefont {Maji},\ and\ \citenamefont
  {Shafi}}]{Lazarides:2023ksx}%
  \BibitemOpen
  \bibfield  {author} {\bibinfo {author} {\bibfnamefont {G.}~\bibnamefont
  {Lazarides}}, \bibinfo {author} {\bibfnamefont {R.}~\bibnamefont {Maji}}, \
  and\ \bibinfo {author} {\bibfnamefont {Q.}~\bibnamefont {Shafi}},\ }\href
  {\doibase 10.1103/PhysRevD.108.095041} {\bibfield  {journal} {\bibinfo
  {journal} {Phys. Rev. D}\ }\textbf {\bibinfo {volume} {108}},\ \bibinfo
  {pages} {095041} (\bibinfo {year} {2023})},\ \Eprint
  {http://arxiv.org/abs/2306.17788} {arXiv:2306.17788 [hep-ph]} \BibitemShut
  {NoStop}%
\bibitem [{\citenamefont {Eichhorn}\ \emph {et~al.}(2024)\citenamefont
  {Eichhorn}, \citenamefont {Lino~dos Santos},\ and\ \citenamefont
  {Miqueleto}}]{Eichhorn:2023gat}%
  \BibitemOpen
  \bibfield  {author} {\bibinfo {author} {\bibfnamefont {A.}~\bibnamefont
  {Eichhorn}}, \bibinfo {author} {\bibfnamefont {R.~R.}\ \bibnamefont {Lino~dos
  Santos}}, \ and\ \bibinfo {author} {\bibfnamefont {J.~a.~L.}\ \bibnamefont
  {Miqueleto}},\ }\href {\doibase 10.1103/PhysRevD.109.026013} {\bibfield
  {journal} {\bibinfo  {journal} {Phys. Rev. D}\ }\textbf {\bibinfo {volume}
  {109}},\ \bibinfo {pages} {026013} (\bibinfo {year} {2024})},\ \Eprint
  {http://arxiv.org/abs/2306.17718} {arXiv:2306.17718 [gr-qc]} \BibitemShut
  {NoStop}%
\bibitem [{\citenamefont {Yamada}\ and\ \citenamefont
  {Yonekura}(2023)}]{Yamada:2023thl}%
  \BibitemOpen
  \bibfield  {author} {\bibinfo {author} {\bibfnamefont {M.}~\bibnamefont
  {Yamada}}\ and\ \bibinfo {author} {\bibfnamefont {K.}~\bibnamefont
  {Yonekura}},\ }\href {\doibase 10.1007/JHEP09(2023)197} {\bibfield  {journal}
  {\bibinfo  {journal} {JHEP}\ }\textbf {\bibinfo {volume} {09}},\ \bibinfo
  {pages} {197} (\bibinfo {year} {2023})},\ \Eprint
  {http://arxiv.org/abs/2307.06586} {arXiv:2307.06586 [hep-ph]} \BibitemShut
  {NoStop}%
\bibitem [{\citenamefont {Franciolini}\ \emph {et~al.}(2023)\citenamefont
  {Franciolini}, \citenamefont {Iovino}, \citenamefont {Vaskonen},\ and\
  \citenamefont {Veermae}}]{Franciolini:2023pbf}%
  \BibitemOpen
  \bibfield  {author} {\bibinfo {author} {\bibfnamefont {G.}~\bibnamefont
  {Franciolini}}, \bibinfo {author} {\bibfnamefont {A.}~\bibnamefont {Iovino},
  \bibfnamefont {Junior.}}, \bibinfo {author} {\bibfnamefont {V.}~\bibnamefont
  {Vaskonen}}, \ and\ \bibinfo {author} {\bibfnamefont {H.}~\bibnamefont
  {Veermae}},\ }\href {\doibase 10.1103/PhysRevLett.131.201401} {\bibfield
  {journal} {\bibinfo  {journal} {Phys. Rev. Lett.}\ }\textbf {\bibinfo
  {volume} {131}},\ \bibinfo {pages} {201401} (\bibinfo {year} {2023})},\
  \Eprint {http://arxiv.org/abs/2306.17149} {arXiv:2306.17149 [astro-ph.CO]}
  \BibitemShut {NoStop}%
\bibitem [{\citenamefont {Vagnozzi}(2023)}]{Vagnozzi:2023lwo}%
  \BibitemOpen
  \bibfield  {author} {\bibinfo {author} {\bibfnamefont {S.}~\bibnamefont
  {Vagnozzi}},\ }\href {\doibase 10.1016/j.jheap.2023.07.001} {\bibfield
  {journal} {\bibinfo  {journal} {JHEAp}\ }\textbf {\bibinfo {volume} {39}},\
  \bibinfo {pages} {81} (\bibinfo {year} {2023})},\ \Eprint
  {http://arxiv.org/abs/2306.16912} {arXiv:2306.16912 [astro-ph.CO]}
  \BibitemShut {NoStop}%
\bibitem [{\citenamefont {Inomata}\ \emph {et~al.}(2024)\citenamefont
  {Inomata}, \citenamefont {Kohri},\ and\ \citenamefont
  {Terada}}]{Inomata:2023zup}%
  \BibitemOpen
  \bibfield  {author} {\bibinfo {author} {\bibfnamefont {K.}~\bibnamefont
  {Inomata}}, \bibinfo {author} {\bibfnamefont {K.}~\bibnamefont {Kohri}}, \
  and\ \bibinfo {author} {\bibfnamefont {T.}~\bibnamefont {Terada}},\ }\href
  {\doibase 10.1103/PhysRevD.109.063506} {\bibfield  {journal} {\bibinfo
  {journal} {Phys. Rev. D}\ }\textbf {\bibinfo {volume} {109}},\ \bibinfo
  {pages} {063506} (\bibinfo {year} {2024})},\ \Eprint
  {http://arxiv.org/abs/2306.17834} {arXiv:2306.17834 [astro-ph.CO]}
  \BibitemShut {NoStop}%
\bibitem [{\citenamefont {Cai}\ \emph {et~al.}(2023)\citenamefont {Cai},
  \citenamefont {He}, \citenamefont {Ma}, \citenamefont {Yan},\ and\
  \citenamefont {Yuan}}]{Cai:2023dls}%
  \BibitemOpen
  \bibfield  {author} {\bibinfo {author} {\bibfnamefont {Y.-F.}\ \bibnamefont
  {Cai}}, \bibinfo {author} {\bibfnamefont {X.-C.}\ \bibnamefont {He}},
  \bibinfo {author} {\bibfnamefont {X.-H.}\ \bibnamefont {Ma}}, \bibinfo
  {author} {\bibfnamefont {S.-F.}\ \bibnamefont {Yan}}, \ and\ \bibinfo
  {author} {\bibfnamefont {G.-W.}\ \bibnamefont {Yuan}},\ }\href {\doibase
  10.1016/j.scib.2023.10.027} {\bibfield  {journal} {\bibinfo  {journal} {Sci.
  Bull.}\ }\textbf {\bibinfo {volume} {68}},\ \bibinfo {pages} {2929} (\bibinfo
  {year} {2023})},\ \Eprint {http://arxiv.org/abs/2306.17822} {arXiv:2306.17822
  [gr-qc]} \BibitemShut {NoStop}%
\bibitem [{\citenamefont {Liu}\ \emph {et~al.}(2024)\citenamefont {Liu},
  \citenamefont {Chen},\ and\ \citenamefont {Huang}}]{Liu:2023ymk}%
  \BibitemOpen
  \bibfield  {author} {\bibinfo {author} {\bibfnamefont {L.}~\bibnamefont
  {Liu}}, \bibinfo {author} {\bibfnamefont {Z.-C.}\ \bibnamefont {Chen}}, \
  and\ \bibinfo {author} {\bibfnamefont {Q.-G.}\ \bibnamefont {Huang}},\ }\href
  {\doibase 10.1103/PhysRevD.109.L061301} {\bibfield  {journal} {\bibinfo
  {journal} {Phys. Rev. D}\ }\textbf {\bibinfo {volume} {109}},\ \bibinfo
  {pages} {L061301} (\bibinfo {year} {2024})},\ \Eprint
  {http://arxiv.org/abs/2307.01102} {arXiv:2307.01102 [astro-ph.CO]}
  \BibitemShut {NoStop}%
\bibitem [{\citenamefont {Abe}\ and\ \citenamefont {Tada}(2023)}]{Abe:2023yrw}%
  \BibitemOpen
  \bibfield  {author} {\bibinfo {author} {\bibfnamefont {K.~T.}\ \bibnamefont
  {Abe}}\ and\ \bibinfo {author} {\bibfnamefont {Y.}~\bibnamefont {Tada}},\
  }\href {\doibase 10.1103/PhysRevD.108.L101304} {\bibfield  {journal}
  {\bibinfo  {journal} {Phys. Rev. D}\ }\textbf {\bibinfo {volume} {108}},\
  \bibinfo {pages} {L101304} (\bibinfo {year} {2023})},\ \Eprint
  {http://arxiv.org/abs/2307.01653} {arXiv:2307.01653 [astro-ph.CO]}
  \BibitemShut {NoStop}%
\bibitem [{\citenamefont {Yi}\ \emph {et~al.}(2023)\citenamefont {Yi},
  \citenamefont {Gao}, \citenamefont {Gong}, \citenamefont {Wang},\ and\
  \citenamefont {Zhang}}]{Yi:2023mbm}%
  \BibitemOpen
  \bibfield  {author} {\bibinfo {author} {\bibfnamefont {Z.}~\bibnamefont
  {Yi}}, \bibinfo {author} {\bibfnamefont {Q.}~\bibnamefont {Gao}}, \bibinfo
  {author} {\bibfnamefont {Y.}~\bibnamefont {Gong}}, \bibinfo {author}
  {\bibfnamefont {Y.}~\bibnamefont {Wang}}, \ and\ \bibinfo {author}
  {\bibfnamefont {F.}~\bibnamefont {Zhang}},\ }\href {\doibase
  10.1007/s11433-023-2266-1} {\bibfield  {journal} {\bibinfo  {journal} {Sci.
  China Phys. Mech. Astron.}\ }\textbf {\bibinfo {volume} {66}},\ \bibinfo
  {pages} {120404} (\bibinfo {year} {2023})},\ \Eprint
  {http://arxiv.org/abs/2307.02467} {arXiv:2307.02467 [gr-qc]} \BibitemShut
  {NoStop}%
\bibitem [{\citenamefont {Firouzjahi}\ and\ \citenamefont
  {Talebian}(2023)}]{Firouzjahi:2023lzg}%
  \BibitemOpen
  \bibfield  {author} {\bibinfo {author} {\bibfnamefont {H.}~\bibnamefont
  {Firouzjahi}}\ and\ \bibinfo {author} {\bibfnamefont {A.}~\bibnamefont
  {Talebian}},\ }\href {\doibase 10.1088/1475-7516/2023/10/032} {\bibfield
  {journal} {\bibinfo  {journal} {JCAP}\ }\textbf {\bibinfo {volume} {10}},\
  \bibinfo {pages} {032} (\bibinfo {year} {2023})},\ \Eprint
  {http://arxiv.org/abs/2307.03164} {arXiv:2307.03164 [gr-qc]} \BibitemShut
  {NoStop}%
\bibitem [{\citenamefont {Salvio}(2023)}]{Salvio:2023ynn}%
  \BibitemOpen
  \bibfield  {author} {\bibinfo {author} {\bibfnamefont {A.}~\bibnamefont
  {Salvio}},\ }\href {\doibase 10.1088/1475-7516/2023/12/046} {\bibfield
  {journal} {\bibinfo  {journal} {JCAP}\ }\textbf {\bibinfo {volume} {12}},\
  \bibinfo {pages} {046} (\bibinfo {year} {2023})},\ \Eprint
  {http://arxiv.org/abs/2307.04694} {arXiv:2307.04694 [hep-ph]} \BibitemShut
  {NoStop}%
\bibitem [{\citenamefont {You}\ \emph {et~al.}(2023)\citenamefont {You},
  \citenamefont {Yi},\ and\ \citenamefont {Wu}}]{You:2023rmn}%
  \BibitemOpen
  \bibfield  {author} {\bibinfo {author} {\bibfnamefont {Z.-Q.}\ \bibnamefont
  {You}}, \bibinfo {author} {\bibfnamefont {Z.}~\bibnamefont {Yi}}, \ and\
  \bibinfo {author} {\bibfnamefont {Y.}~\bibnamefont {Wu}},\ }\href {\doibase
  10.1088/1475-7516/2023/11/065} {\bibfield  {journal} {\bibinfo  {journal}
  {JCAP}\ }\textbf {\bibinfo {volume} {11}},\ \bibinfo {pages} {065} (\bibinfo
  {year} {2023})},\ \Eprint {http://arxiv.org/abs/2307.04419} {arXiv:2307.04419
  [gr-qc]} \BibitemShut {NoStop}%
\bibitem [{\citenamefont {Ye}\ and\ \citenamefont
  {Silvestri}(2024)}]{Ye:2023xyr}%
  \BibitemOpen
  \bibfield  {author} {\bibinfo {author} {\bibfnamefont {G.}~\bibnamefont
  {Ye}}\ and\ \bibinfo {author} {\bibfnamefont {A.}~\bibnamefont {Silvestri}},\
  }\href {\doibase 10.3847/2041-8213/ad2851} {\bibfield  {journal} {\bibinfo
  {journal} {Astrophys. J. Lett.}\ }\textbf {\bibinfo {volume} {963}},\
  \bibinfo {pages} {L15} (\bibinfo {year} {2024})},\ \Eprint
  {http://arxiv.org/abs/2307.05455} {arXiv:2307.05455 [astro-ph.CO]}
  \BibitemShut {NoStop}%
\bibitem [{\citenamefont {Hosseini~Mansoori}\ \emph {et~al.}(2023)\citenamefont
  {Hosseini~Mansoori}, \citenamefont {Felegray}, \citenamefont {Talebian},\
  and\ \citenamefont {Sami}}]{HosseiniMansoori:2023mqh}%
  \BibitemOpen
  \bibfield  {author} {\bibinfo {author} {\bibfnamefont {S.~A.}\ \bibnamefont
  {Hosseini~Mansoori}}, \bibinfo {author} {\bibfnamefont {F.}~\bibnamefont
  {Felegray}}, \bibinfo {author} {\bibfnamefont {A.}~\bibnamefont {Talebian}},
  \ and\ \bibinfo {author} {\bibfnamefont {M.}~\bibnamefont {Sami}},\ }\href
  {\doibase 10.1088/1475-7516/2023/08/067} {\bibfield  {journal} {\bibinfo
  {journal} {JCAP}\ }\textbf {\bibinfo {volume} {08}},\ \bibinfo {pages} {067}
  (\bibinfo {year} {2023})},\ \Eprint {http://arxiv.org/abs/2307.06757}
  {arXiv:2307.06757 [astro-ph.CO]} \BibitemShut {NoStop}%
\bibitem [{\citenamefont {Balaji}\ \emph {et~al.}(2023)\citenamefont {Balaji},
  \citenamefont {Dom\`enech},\ and\ \citenamefont
  {Franciolini}}]{Balaji:2023ehk}%
  \BibitemOpen
  \bibfield  {author} {\bibinfo {author} {\bibfnamefont {S.}~\bibnamefont
  {Balaji}}, \bibinfo {author} {\bibfnamefont {G.}~\bibnamefont {Dom\`enech}},
  \ and\ \bibinfo {author} {\bibfnamefont {G.}~\bibnamefont {Franciolini}},\
  }\href {\doibase 10.1088/1475-7516/2023/10/041} {\bibfield  {journal}
  {\bibinfo  {journal} {JCAP}\ }\textbf {\bibinfo {volume} {10}},\ \bibinfo
  {pages} {041} (\bibinfo {year} {2023})},\ \Eprint
  {http://arxiv.org/abs/2307.08552} {arXiv:2307.08552 [gr-qc]} \BibitemShut
  {NoStop}%
\bibitem [{\citenamefont {Jin}\ \emph {et~al.}(2023)\citenamefont {Jin},
  \citenamefont {Chen}, \citenamefont {Yi}, \citenamefont {You}, \citenamefont
  {Liu},\ and\ \citenamefont {Wu}}]{Jin:2023wri}%
  \BibitemOpen
  \bibfield  {author} {\bibinfo {author} {\bibfnamefont {J.-H.}\ \bibnamefont
  {Jin}}, \bibinfo {author} {\bibfnamefont {Z.-C.}\ \bibnamefont {Chen}},
  \bibinfo {author} {\bibfnamefont {Z.}~\bibnamefont {Yi}}, \bibinfo {author}
  {\bibfnamefont {Z.-Q.}\ \bibnamefont {You}}, \bibinfo {author} {\bibfnamefont
  {L.}~\bibnamefont {Liu}}, \ and\ \bibinfo {author} {\bibfnamefont
  {Y.}~\bibnamefont {Wu}},\ }\href {\doibase 10.1088/1475-7516/2023/09/016}
  {\bibfield  {journal} {\bibinfo  {journal} {JCAP}\ }\textbf {\bibinfo
  {volume} {09}},\ \bibinfo {pages} {016} (\bibinfo {year} {2023})},\ \Eprint
  {http://arxiv.org/abs/2307.08687} {arXiv:2307.08687 [astro-ph.CO]}
  \BibitemShut {NoStop}%
\bibitem [{\citenamefont {Das}\ \emph {et~al.}(2023)\citenamefont {Das},
  \citenamefont {Jaman},\ and\ \citenamefont {Sami}}]{Das:2023nmm}%
  \BibitemOpen
  \bibfield  {author} {\bibinfo {author} {\bibfnamefont {B.}~\bibnamefont
  {Das}}, \bibinfo {author} {\bibfnamefont {N.}~\bibnamefont {Jaman}}, \ and\
  \bibinfo {author} {\bibfnamefont {M.}~\bibnamefont {Sami}},\ }\href {\doibase
  10.1103/PhysRevD.108.103510} {\bibfield  {journal} {\bibinfo  {journal}
  {Phys. Rev. D}\ }\textbf {\bibinfo {volume} {108}},\ \bibinfo {pages}
  {103510} (\bibinfo {year} {2023})},\ \Eprint
  {http://arxiv.org/abs/2307.12913} {arXiv:2307.12913 [gr-qc]} \BibitemShut
  {NoStop}%
\bibitem [{\citenamefont {Ben-Dayan}\ \emph {et~al.}(2023)\citenamefont
  {Ben-Dayan}, \citenamefont {Kumar}, \citenamefont {Thattarampilly},\ and\
  \citenamefont {Verma}}]{Ben-Dayan:2023lwd}%
  \BibitemOpen
  \bibfield  {author} {\bibinfo {author} {\bibfnamefont {I.}~\bibnamefont
  {Ben-Dayan}}, \bibinfo {author} {\bibfnamefont {U.}~\bibnamefont {Kumar}},
  \bibinfo {author} {\bibfnamefont {U.}~\bibnamefont {Thattarampilly}}, \ and\
  \bibinfo {author} {\bibfnamefont {A.}~\bibnamefont {Verma}},\ }\href
  {\doibase 10.1103/PhysRevD.108.103507} {\bibfield  {journal} {\bibinfo
  {journal} {Phys. Rev. D}\ }\textbf {\bibinfo {volume} {108}},\ \bibinfo
  {pages} {103507} (\bibinfo {year} {2023})},\ \Eprint
  {http://arxiv.org/abs/2307.15123} {arXiv:2307.15123 [astro-ph.CO]}
  \BibitemShut {NoStop}%
\bibitem [{\citenamefont {Liu}\ \emph {et~al.}(2023{\natexlab{a}})\citenamefont
  {Liu}, \citenamefont {Chen},\ and\ \citenamefont {Huang}}]{Liu:2023pau}%
  \BibitemOpen
  \bibfield  {author} {\bibinfo {author} {\bibfnamefont {L.}~\bibnamefont
  {Liu}}, \bibinfo {author} {\bibfnamefont {Z.-C.}\ \bibnamefont {Chen}}, \
  and\ \bibinfo {author} {\bibfnamefont {Q.-G.}\ \bibnamefont {Huang}},\ }\href
  {\doibase 10.1088/1475-7516/2023/11/071} {\bibfield  {journal} {\bibinfo
  {journal} {JCAP}\ }\textbf {\bibinfo {volume} {11}},\ \bibinfo {pages} {071}
  (\bibinfo {year} {2023}{\natexlab{a}})},\ \Eprint
  {http://arxiv.org/abs/2307.14911} {arXiv:2307.14911 [astro-ph.CO]}
  \BibitemShut {NoStop}%
\bibitem [{\citenamefont {Yi}\ \emph {et~al.}(2024)\citenamefont {Yi},
  \citenamefont {You},\ and\ \citenamefont {Wu}}]{Yi:2023tdk}%
  \BibitemOpen
  \bibfield  {author} {\bibinfo {author} {\bibfnamefont {Z.}~\bibnamefont
  {Yi}}, \bibinfo {author} {\bibfnamefont {Z.-Q.}\ \bibnamefont {You}}, \ and\
  \bibinfo {author} {\bibfnamefont {Y.}~\bibnamefont {Wu}},\ }\href {\doibase
  10.1088/1475-7516/2024/01/066} {\bibfield  {journal} {\bibinfo  {journal}
  {JCAP}\ }\textbf {\bibinfo {volume} {01}},\ \bibinfo {pages} {066} (\bibinfo
  {year} {2024})},\ \Eprint {http://arxiv.org/abs/2308.05632} {arXiv:2308.05632
  [astro-ph.CO]} \BibitemShut {NoStop}%
\bibitem [{\citenamefont {Frosina}\ and\ \citenamefont
  {Urbano}(2023)}]{Frosina:2023nxu}%
  \BibitemOpen
  \bibfield  {author} {\bibinfo {author} {\bibfnamefont {L.}~\bibnamefont
  {Frosina}}\ and\ \bibinfo {author} {\bibfnamefont {A.}~\bibnamefont
  {Urbano}},\ }\href {\doibase 10.1103/PhysRevD.108.103544} {\bibfield
  {journal} {\bibinfo  {journal} {Phys. Rev. D}\ }\textbf {\bibinfo {volume}
  {108}},\ \bibinfo {pages} {103544} (\bibinfo {year} {2023})},\ \Eprint
  {http://arxiv.org/abs/2308.06915} {arXiv:2308.06915 [astro-ph.CO]}
  \BibitemShut {NoStop}%
\bibitem [{\citenamefont {Bhaumik}\ \emph {et~al.}(2023)\citenamefont
  {Bhaumik}, \citenamefont {Jain},\ and\ \citenamefont
  {Lewicki}}]{Bhaumik:2023wmw}%
  \BibitemOpen
  \bibfield  {author} {\bibinfo {author} {\bibfnamefont {N.}~\bibnamefont
  {Bhaumik}}, \bibinfo {author} {\bibfnamefont {R.~K.}\ \bibnamefont {Jain}}, \
  and\ \bibinfo {author} {\bibfnamefont {M.}~\bibnamefont {Lewicki}},\ }\href
  {\doibase 10.1103/PhysRevD.108.123532} {\bibfield  {journal} {\bibinfo
  {journal} {Phys. Rev. D}\ }\textbf {\bibinfo {volume} {108}},\ \bibinfo
  {pages} {123532} (\bibinfo {year} {2023})},\ \Eprint
  {http://arxiv.org/abs/2308.07912} {arXiv:2308.07912 [astro-ph.CO]}
  \BibitemShut {NoStop}%
\bibitem [{\citenamefont {Yuan}\ \emph {et~al.}(2023)\citenamefont {Yuan},
  \citenamefont {Meng},\ and\ \citenamefont {Huang}}]{Yuan:2023ofl}%
  \BibitemOpen
  \bibfield  {author} {\bibinfo {author} {\bibfnamefont {C.}~\bibnamefont
  {Yuan}}, \bibinfo {author} {\bibfnamefont {D.-S.}\ \bibnamefont {Meng}}, \
  and\ \bibinfo {author} {\bibfnamefont {Q.-G.}\ \bibnamefont {Huang}},\ }\href
  {\doibase 10.1088/1475-7516/2023/12/036} {\bibfield  {journal} {\bibinfo
  {journal} {JCAP}\ }\textbf {\bibinfo {volume} {12}},\ \bibinfo {pages} {036}
  (\bibinfo {year} {2023})},\ \Eprint {http://arxiv.org/abs/2308.07155}
  {arXiv:2308.07155 [astro-ph.CO]} \BibitemShut {NoStop}%
\bibitem [{\citenamefont {Gorji}\ \emph {et~al.}(2023)\citenamefont {Gorji},
  \citenamefont {Sasaki},\ and\ \citenamefont {Suyama}}]{Gorji:2023sil}%
  \BibitemOpen
  \bibfield  {author} {\bibinfo {author} {\bibfnamefont {M.~A.}\ \bibnamefont
  {Gorji}}, \bibinfo {author} {\bibfnamefont {M.}~\bibnamefont {Sasaki}}, \
  and\ \bibinfo {author} {\bibfnamefont {T.}~\bibnamefont {Suyama}},\ }\href
  {\doibase 10.1016/j.physletb.2023.138214} {\bibfield  {journal} {\bibinfo
  {journal} {Phys. Lett. B}\ }\textbf {\bibinfo {volume} {846}},\ \bibinfo
  {pages} {138214} (\bibinfo {year} {2023})},\ \Eprint
  {http://arxiv.org/abs/2307.13109} {arXiv:2307.13109 [astro-ph.CO]}
  \BibitemShut {NoStop}%
\bibitem [{\citenamefont {Afzal}\ \emph {et~al.}(2023)\citenamefont {Afzal}
  \emph {et~al.}}]{NANOGrav:2023hvm}%
  \BibitemOpen
  \bibfield  {author} {\bibinfo {author} {\bibfnamefont {A.}~\bibnamefont
  {Afzal}} \emph {et~al.} (\bibinfo {collaboration} {NANOGrav}),\ }\href
  {\doibase 10.3847/2041-8213/acdc91} {\bibfield  {journal} {\bibinfo
  {journal} {Astrophys. J. Lett.}\ }\textbf {\bibinfo {volume} {951}},\
  \bibinfo {pages} {L11} (\bibinfo {year} {2023})},\ \Eprint
  {http://arxiv.org/abs/2306.16219} {arXiv:2306.16219 [astro-ph.HE]}
  \BibitemShut {NoStop}%
\bibitem [{\citenamefont {Ellis}\ \emph
  {et~al.}(2024{\natexlab{b}})\citenamefont {Ellis}, \citenamefont {Fairbairn},
  \citenamefont {Franciolini}, \citenamefont {H\"utsi}, \citenamefont {Iovino},
  \citenamefont {Lewicki}, \citenamefont {Raidal}, \citenamefont {Urrutia},
  \citenamefont {Vaskonen},\ and\ \citenamefont {Veerm\"ae}}]{Ellis:2023oxs}%
  \BibitemOpen
  \bibfield  {author} {\bibinfo {author} {\bibfnamefont {J.}~\bibnamefont
  {Ellis}}, \bibinfo {author} {\bibfnamefont {M.}~\bibnamefont {Fairbairn}},
  \bibinfo {author} {\bibfnamefont {G.}~\bibnamefont {Franciolini}}, \bibinfo
  {author} {\bibfnamefont {G.}~\bibnamefont {H\"utsi}}, \bibinfo {author}
  {\bibfnamefont {A.}~\bibnamefont {Iovino}}, \bibinfo {author} {\bibfnamefont
  {M.}~\bibnamefont {Lewicki}}, \bibinfo {author} {\bibfnamefont
  {M.}~\bibnamefont {Raidal}}, \bibinfo {author} {\bibfnamefont
  {J.}~\bibnamefont {Urrutia}}, \bibinfo {author} {\bibfnamefont
  {V.}~\bibnamefont {Vaskonen}}, \ and\ \bibinfo {author} {\bibfnamefont
  {H.}~\bibnamefont {Veerm\"ae}},\ }\href {\doibase
  10.1103/PhysRevD.109.023522} {\bibfield  {journal} {\bibinfo  {journal}
  {Phys. Rev. D}\ }\textbf {\bibinfo {volume} {109}},\ \bibinfo {pages}
  {023522} (\bibinfo {year} {2024}{\natexlab{b}})},\ \Eprint
  {http://arxiv.org/abs/2308.08546} {arXiv:2308.08546 [astro-ph.CO]}
  \BibitemShut {NoStop}%
\bibitem [{\citenamefont {Pan}\ and\ \citenamefont {Yang}(2020)}]{Pan:2019uyn}%
  \BibitemOpen
  \bibfield  {author} {\bibinfo {author} {\bibfnamefont {Z.}~\bibnamefont
  {Pan}}\ and\ \bibinfo {author} {\bibfnamefont {H.}~\bibnamefont {Yang}},\
  }\href {\doibase 10.1088/1361-6382/abb074} {\bibfield  {journal} {\bibinfo
  {journal} {Class. Quant. Grav.}\ }\textbf {\bibinfo {volume} {37}},\ \bibinfo
  {pages} {195020} (\bibinfo {year} {2020})},\ \Eprint
  {http://arxiv.org/abs/1910.09637} {arXiv:1910.09637 [astro-ph.CO]}
  \BibitemShut {NoStop}%
\bibitem [{\citenamefont {Biscoveanu}\ \emph {et~al.}(2020)\citenamefont
  {Biscoveanu}, \citenamefont {Talbot}, \citenamefont {Thrane},\ and\
  \citenamefont {Smith}}]{Biscoveanu:2020gds}%
  \BibitemOpen
  \bibfield  {author} {\bibinfo {author} {\bibfnamefont {S.}~\bibnamefont
  {Biscoveanu}}, \bibinfo {author} {\bibfnamefont {C.}~\bibnamefont {Talbot}},
  \bibinfo {author} {\bibfnamefont {E.}~\bibnamefont {Thrane}}, \ and\ \bibinfo
  {author} {\bibfnamefont {R.}~\bibnamefont {Smith}},\ }\href {\doibase
  10.1103/PhysRevLett.125.241101} {\bibfield  {journal} {\bibinfo  {journal}
  {Phys. Rev. Lett.}\ }\textbf {\bibinfo {volume} {125}},\ \bibinfo {pages}
  {241101} (\bibinfo {year} {2020})},\ \Eprint
  {http://arxiv.org/abs/2009.04418} {arXiv:2009.04418 [astro-ph.HE]}
  \BibitemShut {NoStop}%
\bibitem [{\citenamefont {Sharma}\ and\ \citenamefont
  {Harms}(2020)}]{Sharma:2020btq}%
  \BibitemOpen
  \bibfield  {author} {\bibinfo {author} {\bibfnamefont {A.}~\bibnamefont
  {Sharma}}\ and\ \bibinfo {author} {\bibfnamefont {J.}~\bibnamefont {Harms}},\
  }\href {\doibase 10.1103/PhysRevD.102.063009} {\bibfield  {journal} {\bibinfo
   {journal} {Phys. Rev. D}\ }\textbf {\bibinfo {volume} {102}},\ \bibinfo
  {pages} {063009} (\bibinfo {year} {2020})},\ \Eprint
  {http://arxiv.org/abs/2006.16116} {arXiv:2006.16116 [gr-qc]} \BibitemShut
  {NoStop}%
\bibitem [{\citenamefont {Li}\ \emph {et~al.}(2025)\citenamefont {Li},
  \citenamefont {Yu},\ and\ \citenamefont {Pan}}]{Li:2024iua}%
  \BibitemOpen
  \bibfield  {author} {\bibinfo {author} {\bibfnamefont {M.}~\bibnamefont
  {Li}}, \bibinfo {author} {\bibfnamefont {J.}~\bibnamefont {Yu}}, \ and\
  \bibinfo {author} {\bibfnamefont {Z.}~\bibnamefont {Pan}},\ }\href {\doibase
  10.1103/PhysRevD.111.023009} {\bibfield  {journal} {\bibinfo  {journal}
  {Phys. Rev. D}\ }\textbf {\bibinfo {volume} {111}},\ \bibinfo {pages}
  {023009} (\bibinfo {year} {2025})},\ \Eprint
  {http://arxiv.org/abs/2403.01846} {arXiv:2403.01846 [gr-qc]} \BibitemShut
  {NoStop}%
\bibitem [{\citenamefont {Pan}\ and\ \citenamefont {Yang}(2023)}]{Pan:2023naq}%
  \BibitemOpen
  \bibfield  {author} {\bibinfo {author} {\bibfnamefont {Z.}~\bibnamefont
  {Pan}}\ and\ \bibinfo {author} {\bibfnamefont {H.}~\bibnamefont {Yang}},\
  }\href {\doibase 10.1103/PhysRevD.107.123036} {\bibfield  {journal} {\bibinfo
   {journal} {Phys. Rev. D}\ }\textbf {\bibinfo {volume} {107}},\ \bibinfo
  {pages} {123036} (\bibinfo {year} {2023})},\ \Eprint
  {http://arxiv.org/abs/2301.04529} {arXiv:2301.04529 [gr-qc]} \BibitemShut
  {NoStop}%
\bibitem [{\citenamefont {Lewicki}\ and\ \citenamefont
  {Vaskonen}(2023)}]{Lewicki:2021kmu}%
  \BibitemOpen
  \bibfield  {author} {\bibinfo {author} {\bibfnamefont {M.}~\bibnamefont
  {Lewicki}}\ and\ \bibinfo {author} {\bibfnamefont {V.}~\bibnamefont
  {Vaskonen}},\ }\href {\doibase 10.1140/epjc/s10052-023-11323-2} {\bibfield
  {journal} {\bibinfo  {journal} {Eur. Phys. J. C}\ }\textbf {\bibinfo {volume}
  {83}},\ \bibinfo {pages} {168} (\bibinfo {year} {2023})},\ \Eprint
  {http://arxiv.org/abs/2111.05847} {arXiv:2111.05847 [astro-ph.CO]}
  \BibitemShut {NoStop}%
\bibitem [{\citenamefont {Hindmarsh}\ \emph {et~al.}(2015)\citenamefont
  {Hindmarsh}, \citenamefont {Huber}, \citenamefont {Rummukainen},\ and\
  \citenamefont {Weir}}]{Hindmarsh:2015qta}%
  \BibitemOpen
  \bibfield  {author} {\bibinfo {author} {\bibfnamefont {M.}~\bibnamefont
  {Hindmarsh}}, \bibinfo {author} {\bibfnamefont {S.~J.}\ \bibnamefont
  {Huber}}, \bibinfo {author} {\bibfnamefont {K.}~\bibnamefont {Rummukainen}},
  \ and\ \bibinfo {author} {\bibfnamefont {D.~J.}\ \bibnamefont {Weir}},\
  }\href {\doibase 10.1103/PhysRevD.92.123009} {\bibfield  {journal} {\bibinfo
  {journal} {Phys. Rev. D}\ }\textbf {\bibinfo {volume} {92}},\ \bibinfo
  {pages} {123009} (\bibinfo {year} {2015})},\ \Eprint
  {http://arxiv.org/abs/1504.03291} {arXiv:1504.03291 [astro-ph.CO]}
  \BibitemShut {NoStop}%
\bibitem [{\citenamefont {Cutting}\ \emph {et~al.}(2018)\citenamefont
  {Cutting}, \citenamefont {Hindmarsh},\ and\ \citenamefont
  {Weir}}]{Cutting:2018tjt}%
  \BibitemOpen
  \bibfield  {author} {\bibinfo {author} {\bibfnamefont {D.}~\bibnamefont
  {Cutting}}, \bibinfo {author} {\bibfnamefont {M.}~\bibnamefont {Hindmarsh}},
  \ and\ \bibinfo {author} {\bibfnamefont {D.~J.}\ \bibnamefont {Weir}},\
  }\href {\doibase 10.1103/PhysRevD.97.123513} {\bibfield  {journal} {\bibinfo
  {journal} {Phys. Rev. D}\ }\textbf {\bibinfo {volume} {97}},\ \bibinfo
  {pages} {123513} (\bibinfo {year} {2018})},\ \Eprint
  {http://arxiv.org/abs/1802.05712} {arXiv:1802.05712 [astro-ph.CO]}
  \BibitemShut {NoStop}%
\bibitem [{\citenamefont {Di}\ \emph {et~al.}(2021)\citenamefont {Di},
  \citenamefont {Wang}, \citenamefont {Zhou}, \citenamefont {Bian},
  \citenamefont {Cai},\ and\ \citenamefont {Liu}}]{Di:2020kbw}%
  \BibitemOpen
  \bibfield  {author} {\bibinfo {author} {\bibfnamefont {Y.}~\bibnamefont
  {Di}}, \bibinfo {author} {\bibfnamefont {J.}~\bibnamefont {Wang}}, \bibinfo
  {author} {\bibfnamefont {R.}~\bibnamefont {Zhou}}, \bibinfo {author}
  {\bibfnamefont {L.}~\bibnamefont {Bian}}, \bibinfo {author} {\bibfnamefont
  {R.-G.}\ \bibnamefont {Cai}}, \ and\ \bibinfo {author} {\bibfnamefont
  {J.}~\bibnamefont {Liu}},\ }\href {\doibase 10.1103/PhysRevLett.126.251102}
  {\bibfield  {journal} {\bibinfo  {journal} {Phys. Rev. Lett.}\ }\textbf
  {\bibinfo {volume} {126}},\ \bibinfo {pages} {251102} (\bibinfo {year}
  {2021})},\ \Eprint {http://arxiv.org/abs/2012.15625} {arXiv:2012.15625
  [astro-ph.CO]} \BibitemShut {NoStop}%
\bibitem [{\citenamefont {Hiramatsu}\ \emph {et~al.}(2014)\citenamefont
  {Hiramatsu}, \citenamefont {Kawasaki},\ and\ \citenamefont
  {Saikawa}}]{Hiramatsu:2013qaa}%
  \BibitemOpen
  \bibfield  {author} {\bibinfo {author} {\bibfnamefont {T.}~\bibnamefont
  {Hiramatsu}}, \bibinfo {author} {\bibfnamefont {M.}~\bibnamefont {Kawasaki}},
  \ and\ \bibinfo {author} {\bibfnamefont {K.}~\bibnamefont {Saikawa}},\ }\href
  {\doibase 10.1088/1475-7516/2014/02/031} {\bibfield  {journal} {\bibinfo
  {journal} {JCAP}\ }\textbf {\bibinfo {volume} {02}},\ \bibinfo {pages} {031}
  (\bibinfo {year} {2014})},\ \Eprint {http://arxiv.org/abs/1309.5001}
  {arXiv:1309.5001 [astro-ph.CO]} \BibitemShut {NoStop}%
\bibitem [{\citenamefont {Press}\ \emph {et~al.}(1989)\citenamefont {Press},
  \citenamefont {Ryden},\ and\ \citenamefont {Spergel}}]{Press:1989yh}%
  \BibitemOpen
  \bibfield  {author} {\bibinfo {author} {\bibfnamefont {W.~H.}\ \bibnamefont
  {Press}}, \bibinfo {author} {\bibfnamefont {B.~S.}\ \bibnamefont {Ryden}}, \
  and\ \bibinfo {author} {\bibfnamefont {D.~N.}\ \bibnamefont {Spergel}},\
  }\href {\doibase 10.1086/168151} {\bibfield  {journal} {\bibinfo  {journal}
  {Astrophys. J.}\ }\textbf {\bibinfo {volume} {347}},\ \bibinfo {pages} {590}
  (\bibinfo {year} {1989})}\BibitemShut {NoStop}%
\bibitem [{\citenamefont {Liu}(2023)}]{Liu:2023tmv}%
  \BibitemOpen
  \bibfield  {author} {\bibinfo {author} {\bibfnamefont {J.}~\bibnamefont
  {Liu}},\ }\href {\doibase 10.1103/PhysRevD.108.123544} {\bibfield  {journal}
  {\bibinfo  {journal} {Phys. Rev. D}\ }\textbf {\bibinfo {volume} {108}},\
  \bibinfo {pages} {123544} (\bibinfo {year} {2023})}\BibitemShut {NoStop}%
\bibitem [{\citenamefont {Gouttenoire}(2025)}]{Gouttenoire:2025wxc}%
  \BibitemOpen
  \bibfield  {author} {\bibinfo {author} {\bibfnamefont {Y.}~\bibnamefont
  {Gouttenoire}},\ }\href@noop {} {\  (\bibinfo {year} {2025})},\ \Eprint
  {http://arxiv.org/abs/2503.03857} {arXiv:2503.03857 [hep-ph]} \BibitemShut
  {NoStop}%
\bibitem [{\citenamefont {Delos}\ \emph
  {et~al.}(2018{\natexlab{a}})\citenamefont {Delos}, \citenamefont {Erickcek},
  \citenamefont {Bailey},\ and\ \citenamefont {Alvarez}}]{Delos:2017thv}%
  \BibitemOpen
  \bibfield  {author} {\bibinfo {author} {\bibfnamefont {M.~S.}\ \bibnamefont
  {Delos}}, \bibinfo {author} {\bibfnamefont {A.~L.}\ \bibnamefont {Erickcek}},
  \bibinfo {author} {\bibfnamefont {A.~P.}\ \bibnamefont {Bailey}}, \ and\
  \bibinfo {author} {\bibfnamefont {M.~A.}\ \bibnamefont {Alvarez}},\ }\href
  {\doibase 10.1103/PhysRevD.97.041303} {\bibfield  {journal} {\bibinfo
  {journal} {Phys. Rev. D}\ }\textbf {\bibinfo {volume} {97}},\ \bibinfo
  {pages} {041303} (\bibinfo {year} {2018}{\natexlab{a}})},\ \Eprint
  {http://arxiv.org/abs/1712.05421} {arXiv:1712.05421 [astro-ph.CO]}
  \BibitemShut {NoStop}%
\bibitem [{\citenamefont {Delos}\ \emph
  {et~al.}(2018{\natexlab{b}})\citenamefont {Delos}, \citenamefont {Erickcek},
  \citenamefont {Bailey},\ and\ \citenamefont {Alvarez}}]{Delos:2018ueo}%
  \BibitemOpen
  \bibfield  {author} {\bibinfo {author} {\bibfnamefont {M.~S.}\ \bibnamefont
  {Delos}}, \bibinfo {author} {\bibfnamefont {A.~L.}\ \bibnamefont {Erickcek}},
  \bibinfo {author} {\bibfnamefont {A.~P.}\ \bibnamefont {Bailey}}, \ and\
  \bibinfo {author} {\bibfnamefont {M.~A.}\ \bibnamefont {Alvarez}},\ }\href
  {\doibase 10.1103/PhysRevD.98.063527} {\bibfield  {journal} {\bibinfo
  {journal} {Phys. Rev. D}\ }\textbf {\bibinfo {volume} {98}},\ \bibinfo
  {pages} {063527} (\bibinfo {year} {2018}{\natexlab{b}})},\ \Eprint
  {http://arxiv.org/abs/1806.07389} {arXiv:1806.07389 [astro-ph.CO]}
  \BibitemShut {NoStop}%
\bibitem [{\citenamefont {Bringmann}\ \emph {et~al.}(2012)\citenamefont
  {Bringmann}, \citenamefont {Scott},\ and\ \citenamefont
  {Akrami}}]{Bringmann:2011ut}%
  \BibitemOpen
  \bibfield  {author} {\bibinfo {author} {\bibfnamefont {T.}~\bibnamefont
  {Bringmann}}, \bibinfo {author} {\bibfnamefont {P.}~\bibnamefont {Scott}}, \
  and\ \bibinfo {author} {\bibfnamefont {Y.}~\bibnamefont {Akrami}},\ }\href
  {\doibase 10.1103/PhysRevD.85.125027} {\bibfield  {journal} {\bibinfo
  {journal} {Phys. Rev. D}\ }\textbf {\bibinfo {volume} {85}},\ \bibinfo
  {pages} {125027} (\bibinfo {year} {2012})},\ \Eprint
  {http://arxiv.org/abs/1110.2484} {arXiv:1110.2484 [astro-ph.CO]} \BibitemShut
  {NoStop}%
\bibitem [{\citenamefont {Rosado}\ \emph {et~al.}(2015)\citenamefont {Rosado},
  \citenamefont {Sesana},\ and\ \citenamefont {Gair}}]{Rosado:2015epa}%
  \BibitemOpen
  \bibfield  {author} {\bibinfo {author} {\bibfnamefont {P.~A.}\ \bibnamefont
  {Rosado}}, \bibinfo {author} {\bibfnamefont {A.}~\bibnamefont {Sesana}}, \
  and\ \bibinfo {author} {\bibfnamefont {J.}~\bibnamefont {Gair}},\ }\href
  {\doibase 10.1093/mnras/stv1098} {\bibfield  {journal} {\bibinfo  {journal}
  {Mon. Not. Roy. Astron. Soc.}\ }\textbf {\bibinfo {volume} {451}},\ \bibinfo
  {pages} {2417} (\bibinfo {year} {2015})},\ \Eprint
  {http://arxiv.org/abs/1503.04803} {arXiv:1503.04803 [astro-ph.HE]}
  \BibitemShut {NoStop}%
\bibitem [{\citenamefont {Punturo}\ \emph {et~al.}(2010)\citenamefont {Punturo}
  \emph {et~al.}}]{Punturo:2010zz}%
  \BibitemOpen
  \bibfield  {author} {\bibinfo {author} {\bibfnamefont {M.}~\bibnamefont
  {Punturo}} \emph {et~al.},\ }\href {\doibase 10.1088/0264-9381/27/19/194002}
  {\bibfield  {journal} {\bibinfo  {journal} {Class. Quant. Grav.}\ }\textbf
  {\bibinfo {volume} {27}},\ \bibinfo {pages} {194002} (\bibinfo {year}
  {2010})}\BibitemShut {NoStop}%
\bibitem [{\citenamefont {Reitze}\ \emph {et~al.}(2019)\citenamefont {Reitze}
  \emph {et~al.}}]{Reitze:2019iox}%
  \BibitemOpen
  \bibfield  {author} {\bibinfo {author} {\bibfnamefont {D.}~\bibnamefont
  {Reitze}} \emph {et~al.},\ }\href@noop {} {\bibfield  {journal} {\bibinfo
  {journal} {Bull. Am. Astron. Soc.}\ }\textbf {\bibinfo {volume} {51}},\
  \bibinfo {pages} {035} (\bibinfo {year} {2019})},\ \Eprint
  {http://arxiv.org/abs/1907.04833} {arXiv:1907.04833 [astro-ph.IM]}
  \BibitemShut {NoStop}%
\bibitem [{\citenamefont {Caprini}\ \emph {et~al.}(2016)\citenamefont {Caprini}
  \emph {et~al.}}]{Caprini:2015zlo}%
  \BibitemOpen
  \bibfield  {author} {\bibinfo {author} {\bibfnamefont {C.}~\bibnamefont
  {Caprini}} \emph {et~al.},\ }\href {\doibase 10.1088/1475-7516/2016/04/001}
  {\bibfield  {journal} {\bibinfo  {journal} {JCAP}\ }\textbf {\bibinfo
  {volume} {04}},\ \bibinfo {pages} {001} (\bibinfo {year} {2016})},\ \Eprint
  {http://arxiv.org/abs/1512.06239} {arXiv:1512.06239 [astro-ph.CO]}
  \BibitemShut {NoStop}%
\bibitem [{\citenamefont {Caprini}\ \emph {et~al.}(2020)\citenamefont {Caprini}
  \emph {et~al.}}]{Caprini:2019egz}%
  \BibitemOpen
  \bibfield  {author} {\bibinfo {author} {\bibfnamefont {C.}~\bibnamefont
  {Caprini}} \emph {et~al.},\ }\href {\doibase 10.1088/1475-7516/2020/03/024}
  {\bibfield  {journal} {\bibinfo  {journal} {JCAP}\ }\textbf {\bibinfo
  {volume} {03}},\ \bibinfo {pages} {024} (\bibinfo {year} {2020})},\ \Eprint
  {http://arxiv.org/abs/1910.13125} {arXiv:1910.13125 [astro-ph.CO]}
  \BibitemShut {NoStop}%
\bibitem [{\citenamefont {Weir}(2018)}]{Weir:2017wfa}%
  \BibitemOpen
  \bibfield  {author} {\bibinfo {author} {\bibfnamefont {D.~J.}\ \bibnamefont
  {Weir}},\ }\href {\doibase 10.1098/rsta.2017.0126} {\bibfield  {journal}
  {\bibinfo  {journal} {Phil. Trans. Roy. Soc. Lond. A}\ }\textbf {\bibinfo
  {volume} {376}},\ \bibinfo {pages} {20170126} (\bibinfo {year} {2018})},\
  \bibinfo {note} {[Erratum: Phil.Trans.Roy.Soc.Lond.A 381, 20230212 (2023)]},\
  \Eprint {http://arxiv.org/abs/1705.01783} {arXiv:1705.01783 [hep-ph]}
  \BibitemShut {NoStop}%
\bibitem [{\citenamefont {Mazumdar}\ and\ \citenamefont
  {White}(2019)}]{Mazumdar:2018dfl}%
  \BibitemOpen
  \bibfield  {author} {\bibinfo {author} {\bibfnamefont {A.}~\bibnamefont
  {Mazumdar}}\ and\ \bibinfo {author} {\bibfnamefont {G.}~\bibnamefont
  {White}},\ }\href {\doibase 10.1088/1361-6633/ab1f55} {\bibfield  {journal}
  {\bibinfo  {journal} {Rept. Prog. Phys.}\ }\textbf {\bibinfo {volume} {82}},\
  \bibinfo {pages} {076901} (\bibinfo {year} {2019})},\ \Eprint
  {http://arxiv.org/abs/1811.01948} {arXiv:1811.01948 [hep-ph]} \BibitemShut
  {NoStop}%
\bibitem [{\citenamefont {Liu}\ \emph {et~al.}(2023{\natexlab{b}})\citenamefont
  {Liu}, \citenamefont {Bian}, \citenamefont {Cai}, \citenamefont {Guo},\ and\
  \citenamefont {Wang}}]{Liu:2022lvz}%
  \BibitemOpen
  \bibfield  {author} {\bibinfo {author} {\bibfnamefont {J.}~\bibnamefont
  {Liu}}, \bibinfo {author} {\bibfnamefont {L.}~\bibnamefont {Bian}}, \bibinfo
  {author} {\bibfnamefont {R.-G.}\ \bibnamefont {Cai}}, \bibinfo {author}
  {\bibfnamefont {Z.-K.}\ \bibnamefont {Guo}}, \ and\ \bibinfo {author}
  {\bibfnamefont {S.-J.}\ \bibnamefont {Wang}},\ }\href {\doibase
  10.1103/PhysRevLett.130.051001} {\bibfield  {journal} {\bibinfo  {journal}
  {Phys. Rev. Lett.}\ }\textbf {\bibinfo {volume} {130}},\ \bibinfo {pages}
  {051001} (\bibinfo {year} {2023}{\natexlab{b}})},\ \Eprint
  {http://arxiv.org/abs/2208.14086} {arXiv:2208.14086 [astro-ph.CO]}
  \BibitemShut {NoStop}%
\bibitem [{\citenamefont {Cai}\ \emph {et~al.}(2018)\citenamefont {Cai},
  \citenamefont {Tong}, \citenamefont {Wang},\ and\ \citenamefont
  {Yan}}]{Cai:2018tuh}%
  \BibitemOpen
  \bibfield  {author} {\bibinfo {author} {\bibfnamefont {Y.-F.}\ \bibnamefont
  {Cai}}, \bibinfo {author} {\bibfnamefont {X.}~\bibnamefont {Tong}}, \bibinfo
  {author} {\bibfnamefont {D.-G.}\ \bibnamefont {Wang}}, \ and\ \bibinfo
  {author} {\bibfnamefont {S.-F.}\ \bibnamefont {Yan}},\ }\href {\doibase
  10.1103/PhysRevLett.121.081306} {\bibfield  {journal} {\bibinfo  {journal}
  {Phys. Rev. Lett.}\ }\textbf {\bibinfo {volume} {121}},\ \bibinfo {pages}
  {081306} (\bibinfo {year} {2018})},\ \Eprint
  {http://arxiv.org/abs/1805.03639} {arXiv:1805.03639 [astro-ph.CO]}
  \BibitemShut {NoStop}%
\bibitem [{\citenamefont {Cai}\ \emph {et~al.}(2020)\citenamefont {Cai},
  \citenamefont {Guo}, \citenamefont {Liu}, \citenamefont {Liu},\ and\
  \citenamefont {Yang}}]{Cai:2019bmk}%
  \BibitemOpen
  \bibfield  {author} {\bibinfo {author} {\bibfnamefont {R.-G.}\ \bibnamefont
  {Cai}}, \bibinfo {author} {\bibfnamefont {Z.-K.}\ \bibnamefont {Guo}},
  \bibinfo {author} {\bibfnamefont {J.}~\bibnamefont {Liu}}, \bibinfo {author}
  {\bibfnamefont {L.}~\bibnamefont {Liu}}, \ and\ \bibinfo {author}
  {\bibfnamefont {X.-Y.}\ \bibnamefont {Yang}},\ }\href {\doibase
  10.1088/1475-7516/2020/06/013} {\bibfield  {journal} {\bibinfo  {journal}
  {JCAP}\ }\textbf {\bibinfo {volume} {06}},\ \bibinfo {pages} {013} (\bibinfo
  {year} {2020})},\ \Eprint {http://arxiv.org/abs/1912.10437} {arXiv:1912.10437
  [astro-ph.CO]} \BibitemShut {NoStop}%
\bibitem [{\citenamefont {Saikawa}(2017)}]{Saikawa:2017hiv}%
  \BibitemOpen
  \bibfield  {author} {\bibinfo {author} {\bibfnamefont {K.}~\bibnamefont
  {Saikawa}},\ }\href {\doibase 10.3390/universe3020040} {\bibfield  {journal}
  {\bibinfo  {journal} {Universe}\ }\textbf {\bibinfo {volume} {3}},\ \bibinfo
  {pages} {40} (\bibinfo {year} {2017})},\ \Eprint
  {http://arxiv.org/abs/1703.02576} {arXiv:1703.02576 [hep-ph]} \BibitemShut
  {NoStop}%
\bibitem [{\citenamefont {Sesana}\ \emph {et~al.}(2008)\citenamefont {Sesana},
  \citenamefont {Vecchio},\ and\ \citenamefont {Colacino}}]{Sesana:2008mz}%
  \BibitemOpen
  \bibfield  {author} {\bibinfo {author} {\bibfnamefont {A.}~\bibnamefont
  {Sesana}}, \bibinfo {author} {\bibfnamefont {A.}~\bibnamefont {Vecchio}}, \
  and\ \bibinfo {author} {\bibfnamefont {C.~N.}\ \bibnamefont {Colacino}},\
  }\href {\doibase 10.1111/j.1365-2966.2008.13682.x} {\bibfield  {journal}
  {\bibinfo  {journal} {Mon. Not. Roy. Astron. Soc.}\ }\textbf {\bibinfo
  {volume} {390}},\ \bibinfo {pages} {192} (\bibinfo {year} {2008})},\ \Eprint
  {http://arxiv.org/abs/0804.4476} {arXiv:0804.4476 [astro-ph]} \BibitemShut
  {NoStop}%
\bibitem [{\citenamefont {Janssen}\ \emph {et~al.}(2015)\citenamefont {Janssen}
  \emph {et~al.}}]{Janssen:2014dka}%
  \BibitemOpen
  \bibfield  {author} {\bibinfo {author} {\bibfnamefont {G.}~\bibnamefont
  {Janssen}} \emph {et~al.},\ }\href {\doibase 10.22323/1.215.0037} {\bibfield
  {journal} {\bibinfo  {journal} {PoS}\ }\textbf {\bibinfo {volume}
  {AASKA14}},\ \bibinfo {pages} {037} (\bibinfo {year} {2015})},\ \Eprint
  {http://arxiv.org/abs/1501.00127} {arXiv:1501.00127 [astro-ph.IM]}
  \BibitemShut {NoStop}%
\bibitem [{\citenamefont {\"Ozsoy}\ and\ \citenamefont
  {Tasinato}(2023)}]{Ozsoy:2023ryl}%
  \BibitemOpen
  \bibfield  {author} {\bibinfo {author} {\bibfnamefont {O.}~\bibnamefont
  {\"Ozsoy}}\ and\ \bibinfo {author} {\bibfnamefont {G.}~\bibnamefont
  {Tasinato}},\ }\href {\doibase 10.3390/universe9050203} {\  (\bibinfo {year}
  {2023}),\ 10.3390/universe9050203},\ \Eprint
  {http://arxiv.org/abs/2301.03600} {arXiv:2301.03600 [astro-ph.CO]}
  \BibitemShut {NoStop}%
\bibitem [{\citenamefont {Press}\ and\ \citenamefont
  {Schechter}(1974)}]{Press:1973iz}%
  \BibitemOpen
  \bibfield  {author} {\bibinfo {author} {\bibfnamefont {W.~H.}\ \bibnamefont
  {Press}}\ and\ \bibinfo {author} {\bibfnamefont {P.}~\bibnamefont
  {Schechter}},\ }\href {\doibase 10.1086/152650} {\bibfield  {journal}
  {\bibinfo  {journal} {Astrophys. J.}\ }\textbf {\bibinfo {volume} {187}},\
  \bibinfo {pages} {425} (\bibinfo {year} {1974})}\BibitemShut {NoStop}%
\bibitem [{\citenamefont {Kohri}\ and\ \citenamefont
  {Terada}(2018)}]{Kohri:2018awv}%
  \BibitemOpen
  \bibfield  {author} {\bibinfo {author} {\bibfnamefont {K.}~\bibnamefont
  {Kohri}}\ and\ \bibinfo {author} {\bibfnamefont {T.}~\bibnamefont {Terada}},\
  }\href {\doibase 10.1103/PhysRevD.97.123532} {\bibfield  {journal} {\bibinfo
  {journal} {Phys. Rev. D}\ }\textbf {\bibinfo {volume} {97}},\ \bibinfo
  {pages} {123532} (\bibinfo {year} {2018})},\ \Eprint
  {http://arxiv.org/abs/1804.08577} {arXiv:1804.08577 [gr-qc]} \BibitemShut
  {NoStop}%
\bibitem [{\citenamefont {Espinosa}\ \emph {et~al.}(2018)\citenamefont
  {Espinosa}, \citenamefont {Racco},\ and\ \citenamefont
  {Riotto}}]{Espinosa:2018eve}%
  \BibitemOpen
  \bibfield  {author} {\bibinfo {author} {\bibfnamefont {J.~R.}\ \bibnamefont
  {Espinosa}}, \bibinfo {author} {\bibfnamefont {D.}~\bibnamefont {Racco}}, \
  and\ \bibinfo {author} {\bibfnamefont {A.}~\bibnamefont {Riotto}},\ }\href
  {\doibase 10.1088/1475-7516/2018/09/012} {\bibfield  {journal} {\bibinfo
  {journal} {JCAP}\ }\textbf {\bibinfo {volume} {09}},\ \bibinfo {pages} {012}
  (\bibinfo {year} {2018})},\ \Eprint {http://arxiv.org/abs/1804.07732}
  {arXiv:1804.07732 [hep-ph]} \BibitemShut {NoStop}%
\bibitem [{\citenamefont {Delos}\ and\ \citenamefont
  {Franciolini}(2023)}]{Delos:2023fpm}%
  \BibitemOpen
  \bibfield  {author} {\bibinfo {author} {\bibfnamefont {M.~S.}\ \bibnamefont
  {Delos}}\ and\ \bibinfo {author} {\bibfnamefont {G.}~\bibnamefont
  {Franciolini}},\ }\href {\doibase 10.1103/PhysRevD.107.083505} {\bibfield
  {journal} {\bibinfo  {journal} {Phys. Rev. D}\ }\textbf {\bibinfo {volume}
  {107}},\ \bibinfo {pages} {083505} (\bibinfo {year} {2023})},\ \Eprint
  {http://arxiv.org/abs/2301.13171} {arXiv:2301.13171 [astro-ph.CO]}
  \BibitemShut {NoStop}%
\bibitem [{\citenamefont {Sten~Delos}\ and\ \citenamefont
  {Silk}(2023)}]{StenDelos:2022jld}%
  \BibitemOpen
  \bibfield  {author} {\bibinfo {author} {\bibfnamefont {M.}~\bibnamefont
  {Sten~Delos}}\ and\ \bibinfo {author} {\bibfnamefont {J.}~\bibnamefont
  {Silk}},\ }\href {\doibase 10.1093/mnras/stad356} {\bibfield  {journal}
  {\bibinfo  {journal} {Mon. Not. Roy. Astron. Soc.}\ }\textbf {\bibinfo
  {volume} {520}},\ \bibinfo {pages} {4370} (\bibinfo {year} {2023})},\ \Eprint
  {http://arxiv.org/abs/2210.04904} {arXiv:2210.04904 [astro-ph.CO]}
  \BibitemShut {NoStop}%
\bibitem [{\citenamefont {Barnaby}(2010)}]{Barnaby:2010ke}%
  \BibitemOpen
  \bibfield  {author} {\bibinfo {author} {\bibfnamefont {N.}~\bibnamefont
  {Barnaby}},\ }\href {\doibase 10.1103/PhysRevD.82.106009} {\bibfield
  {journal} {\bibinfo  {journal} {Phys. Rev. D}\ }\textbf {\bibinfo {volume}
  {82}},\ \bibinfo {pages} {106009} (\bibinfo {year} {2010})},\ \Eprint
  {http://arxiv.org/abs/1006.4615} {arXiv:1006.4615 [astro-ph.CO]} \BibitemShut
  {NoStop}%
\bibitem [{\citenamefont {Cai}\ \emph {et~al.}(2021)\citenamefont {Cai},
  \citenamefont {Chen},\ and\ \citenamefont {Fu}}]{Cai:2021wzd}%
  \BibitemOpen
  \bibfield  {author} {\bibinfo {author} {\bibfnamefont {R.-G.}\ \bibnamefont
  {Cai}}, \bibinfo {author} {\bibfnamefont {C.}~\bibnamefont {Chen}}, \ and\
  \bibinfo {author} {\bibfnamefont {C.}~\bibnamefont {Fu}},\ }\href {\doibase
  10.1103/PhysRevD.104.083537} {\bibfield  {journal} {\bibinfo  {journal}
  {Phys. Rev. D}\ }\textbf {\bibinfo {volume} {104}},\ \bibinfo {pages}
  {083537} (\bibinfo {year} {2021})},\ \Eprint
  {http://arxiv.org/abs/2108.03422} {arXiv:2108.03422 [astro-ph.CO]}
  \BibitemShut {NoStop}%
\bibitem [{\citenamefont {Arzoumanian}\ \emph {et~al.}(2021)\citenamefont
  {Arzoumanian} \emph {et~al.}}]{NANOGrav:2021flc}%
  \BibitemOpen
  \bibfield  {author} {\bibinfo {author} {\bibfnamefont {Z.}~\bibnamefont
  {Arzoumanian}} \emph {et~al.} (\bibinfo {collaboration} {NANOGrav}),\ }\href
  {\doibase 10.1103/PhysRevLett.127.251302} {\bibfield  {journal} {\bibinfo
  {journal} {Phys. Rev. Lett.}\ }\textbf {\bibinfo {volume} {127}},\ \bibinfo
  {pages} {251302} (\bibinfo {year} {2021})},\ \Eprint
  {http://arxiv.org/abs/2104.13930} {arXiv:2104.13930 [astro-ph.CO]}
  \BibitemShut {NoStop}%
\bibitem [{\citenamefont {Espinosa}\ \emph {et~al.}(2010)\citenamefont
  {Espinosa}, \citenamefont {Konstandin}, \citenamefont {No},\ and\
  \citenamefont {Servant}}]{Espinosa:2010hh}%
  \BibitemOpen
  \bibfield  {author} {\bibinfo {author} {\bibfnamefont {J.~R.}\ \bibnamefont
  {Espinosa}}, \bibinfo {author} {\bibfnamefont {T.}~\bibnamefont
  {Konstandin}}, \bibinfo {author} {\bibfnamefont {J.~M.}\ \bibnamefont {No}},
  \ and\ \bibinfo {author} {\bibfnamefont {G.}~\bibnamefont {Servant}},\ }\href
  {\doibase 10.1088/1475-7516/2010/06/028} {\bibfield  {journal} {\bibinfo
  {journal} {JCAP}\ }\textbf {\bibinfo {volume} {06}},\ \bibinfo {pages} {028}
  (\bibinfo {year} {2010})},\ \Eprint {http://arxiv.org/abs/1004.4187}
  {arXiv:1004.4187 [hep-ph]} \BibitemShut {NoStop}%
\bibitem [{\citenamefont {Kofman}\ \emph {et~al.}(1997)\citenamefont {Kofman},
  \citenamefont {Linde},\ and\ \citenamefont {Starobinsky}}]{Kofman:1997yn}%
  \BibitemOpen
  \bibfield  {author} {\bibinfo {author} {\bibfnamefont {L.}~\bibnamefont
  {Kofman}}, \bibinfo {author} {\bibfnamefont {A.~D.}\ \bibnamefont {Linde}}, \
  and\ \bibinfo {author} {\bibfnamefont {A.~A.}\ \bibnamefont {Starobinsky}},\
  }\href {\doibase 10.1103/PhysRevD.56.3258} {\bibfield  {journal} {\bibinfo
  {journal} {Phys. Rev. D}\ }\textbf {\bibinfo {volume} {56}},\ \bibinfo
  {pages} {3258} (\bibinfo {year} {1997})},\ \Eprint
  {http://arxiv.org/abs/hep-ph/9704452} {arXiv:hep-ph/9704452} \BibitemShut
  {NoStop}%
\bibitem [{\citenamefont {Khlebnikov}\ and\ \citenamefont
  {Tkachev}(1997)}]{Khlebnikov:1997di}%
  \BibitemOpen
  \bibfield  {author} {\bibinfo {author} {\bibfnamefont {S.~Y.}\ \bibnamefont
  {Khlebnikov}}\ and\ \bibinfo {author} {\bibfnamefont {I.~I.}\ \bibnamefont
  {Tkachev}},\ }\href {\doibase 10.1103/PhysRevD.56.653} {\bibfield  {journal}
  {\bibinfo  {journal} {Phys. Rev. D}\ }\textbf {\bibinfo {volume} {56}},\
  \bibinfo {pages} {653} (\bibinfo {year} {1997})},\ \Eprint
  {http://arxiv.org/abs/hep-ph/9701423} {arXiv:hep-ph/9701423} \BibitemShut
  {NoStop}%
\bibitem [{\citenamefont {Amin}\ \emph {et~al.}(2010)\citenamefont {Amin},
  \citenamefont {Easther},\ and\ \citenamefont {Finkel}}]{Amin:2010dc}%
  \BibitemOpen
  \bibfield  {author} {\bibinfo {author} {\bibfnamefont {M.~A.}\ \bibnamefont
  {Amin}}, \bibinfo {author} {\bibfnamefont {R.}~\bibnamefont {Easther}}, \
  and\ \bibinfo {author} {\bibfnamefont {H.}~\bibnamefont {Finkel}},\ }\href
  {\doibase 10.1088/1475-7516/2010/12/001} {\bibfield  {journal} {\bibinfo
  {journal} {JCAP}\ }\textbf {\bibinfo {volume} {12}},\ \bibinfo {pages} {001}
  (\bibinfo {year} {2010})},\ \Eprint {http://arxiv.org/abs/1009.2505}
  {arXiv:1009.2505 [astro-ph.CO]} \BibitemShut {NoStop}%
\bibitem [{\citenamefont {Zhou}\ \emph {et~al.}(2013)\citenamefont {Zhou},
  \citenamefont {Copeland}, \citenamefont {Easther}, \citenamefont {Finkel},
  \citenamefont {Mou},\ and\ \citenamefont {Saffin}}]{Zhou:2013tsa}%
  \BibitemOpen
  \bibfield  {author} {\bibinfo {author} {\bibfnamefont {S.-Y.}\ \bibnamefont
  {Zhou}}, \bibinfo {author} {\bibfnamefont {E.~J.}\ \bibnamefont {Copeland}},
  \bibinfo {author} {\bibfnamefont {R.}~\bibnamefont {Easther}}, \bibinfo
  {author} {\bibfnamefont {H.}~\bibnamefont {Finkel}}, \bibinfo {author}
  {\bibfnamefont {Z.-G.}\ \bibnamefont {Mou}}, \ and\ \bibinfo {author}
  {\bibfnamefont {P.~M.}\ \bibnamefont {Saffin}},\ }\href {\doibase
  10.1007/JHEP10(2013)026} {\bibfield  {journal} {\bibinfo  {journal} {JHEP}\
  }\textbf {\bibinfo {volume} {10}},\ \bibinfo {pages} {026} (\bibinfo {year}
  {2013})},\ \Eprint {http://arxiv.org/abs/1304.6094} {arXiv:1304.6094
  [astro-ph.CO]} \BibitemShut {NoStop}%
\bibitem [{\citenamefont {Lozanov}\ and\ \citenamefont
  {Amin}(2019)}]{Lozanov:2019ylm}%
  \BibitemOpen
  \bibfield  {author} {\bibinfo {author} {\bibfnamefont {K.~D.}\ \bibnamefont
  {Lozanov}}\ and\ \bibinfo {author} {\bibfnamefont {M.~A.}\ \bibnamefont
  {Amin}},\ }\href {\doibase 10.1103/PhysRevD.99.123504} {\bibfield  {journal}
  {\bibinfo  {journal} {Phys. Rev. D}\ }\textbf {\bibinfo {volume} {99}},\
  \bibinfo {pages} {123504} (\bibinfo {year} {2019})},\ \Eprint
  {http://arxiv.org/abs/1902.06736} {arXiv:1902.06736 [astro-ph.CO]}
  \BibitemShut {NoStop}%
\bibitem [{\citenamefont {Liu}\ \emph {et~al.}(2018)\citenamefont {Liu},
  \citenamefont {Guo}, \citenamefont {Cai},\ and\ \citenamefont
  {Shiu}}]{Liu:2017hua}%
  \BibitemOpen
  \bibfield  {author} {\bibinfo {author} {\bibfnamefont {J.}~\bibnamefont
  {Liu}}, \bibinfo {author} {\bibfnamefont {Z.-K.}\ \bibnamefont {Guo}},
  \bibinfo {author} {\bibfnamefont {R.-G.}\ \bibnamefont {Cai}}, \ and\
  \bibinfo {author} {\bibfnamefont {G.}~\bibnamefont {Shiu}},\ }\href {\doibase
  10.1103/PhysRevLett.120.031301} {\bibfield  {journal} {\bibinfo  {journal}
  {Phys. Rev. Lett.}\ }\textbf {\bibinfo {volume} {120}},\ \bibinfo {pages}
  {031301} (\bibinfo {year} {2018})},\ \Eprint
  {http://arxiv.org/abs/1707.09841} {arXiv:1707.09841 [astro-ph.CO]}
  \BibitemShut {NoStop}%
\bibitem [{\citenamefont {Oll\'e}\ \emph {et~al.}(2020)\citenamefont {Oll\'e},
  \citenamefont {Pujol\`as},\ and\ \citenamefont {Rompineve}}]{Olle:2019kbo}%
  \BibitemOpen
  \bibfield  {author} {\bibinfo {author} {\bibfnamefont {J.}~\bibnamefont
  {Oll\'e}}, \bibinfo {author} {\bibfnamefont {O.}~\bibnamefont {Pujol\`as}}, \
  and\ \bibinfo {author} {\bibfnamefont {F.}~\bibnamefont {Rompineve}},\ }\href
  {\doibase 10.1088/1475-7516/2020/02/006} {\bibfield  {journal} {\bibinfo
  {journal} {JCAP}\ }\textbf {\bibinfo {volume} {02}},\ \bibinfo {pages} {006}
  (\bibinfo {year} {2020})},\ \Eprint {http://arxiv.org/abs/1906.06352}
  {arXiv:1906.06352 [hep-ph]} \BibitemShut {NoStop}%
\bibitem [{\citenamefont {Inomata}\ \emph {et~al.}(2021)\citenamefont
  {Inomata}, \citenamefont {Kawasaki}, \citenamefont {Mukaida},\ and\
  \citenamefont {Yanagida}}]{Inomata:2020xad}%
  \BibitemOpen
  \bibfield  {author} {\bibinfo {author} {\bibfnamefont {K.}~\bibnamefont
  {Inomata}}, \bibinfo {author} {\bibfnamefont {M.}~\bibnamefont {Kawasaki}},
  \bibinfo {author} {\bibfnamefont {K.}~\bibnamefont {Mukaida}}, \ and\
  \bibinfo {author} {\bibfnamefont {T.~T.}\ \bibnamefont {Yanagida}},\ }\href
  {\doibase 10.1103/PhysRevLett.126.131301} {\bibfield  {journal} {\bibinfo
  {journal} {Phys. Rev. Lett.}\ }\textbf {\bibinfo {volume} {126}},\ \bibinfo
  {pages} {131301} (\bibinfo {year} {2021})},\ \Eprint
  {http://arxiv.org/abs/2011.01270} {arXiv:2011.01270 [astro-ph.CO]}
  \BibitemShut {NoStop}%
\bibitem [{\citenamefont {Kitajima}\ \emph {et~al.}(2021)\citenamefont
  {Kitajima}, \citenamefont {Soda},\ and\ \citenamefont
  {Urakawa}}]{Kitajima:2020rpm}%
  \BibitemOpen
  \bibfield  {author} {\bibinfo {author} {\bibfnamefont {N.}~\bibnamefont
  {Kitajima}}, \bibinfo {author} {\bibfnamefont {J.}~\bibnamefont {Soda}}, \
  and\ \bibinfo {author} {\bibfnamefont {Y.}~\bibnamefont {Urakawa}},\ }\href
  {\doibase 10.1103/PhysRevLett.126.121301} {\bibfield  {journal} {\bibinfo
  {journal} {Phys. Rev. Lett.}\ }\textbf {\bibinfo {volume} {126}},\ \bibinfo
  {pages} {121301} (\bibinfo {year} {2021})},\ \Eprint
  {http://arxiv.org/abs/2010.10990} {arXiv:2010.10990 [astro-ph.CO]}
  \BibitemShut {NoStop}%
\bibitem [{\citenamefont {Ramberg}\ and\ \citenamefont
  {Visinelli}(2021)}]{Ramberg:2020oct}%
  \BibitemOpen
  \bibfield  {author} {\bibinfo {author} {\bibfnamefont {N.}~\bibnamefont
  {Ramberg}}\ and\ \bibinfo {author} {\bibfnamefont {L.}~\bibnamefont
  {Visinelli}},\ }\href {\doibase 10.1103/PhysRevD.103.063031} {\bibfield
  {journal} {\bibinfo  {journal} {Phys. Rev. D}\ }\textbf {\bibinfo {volume}
  {103}},\ \bibinfo {pages} {063031} (\bibinfo {year} {2021})},\ \Eprint
  {http://arxiv.org/abs/2012.06882} {arXiv:2012.06882 [astro-ph.CO]}
  \BibitemShut {NoStop}%
\bibitem [{\citenamefont {Kawasaki}\ and\ \citenamefont
  {Nakatsuka}(2021)}]{Kawasaki:2021ycf}%
  \BibitemOpen
  \bibfield  {author} {\bibinfo {author} {\bibfnamefont {M.}~\bibnamefont
  {Kawasaki}}\ and\ \bibinfo {author} {\bibfnamefont {H.}~\bibnamefont
  {Nakatsuka}},\ }\href {\doibase 10.1088/1475-7516/2021/05/023} {\bibfield
  {journal} {\bibinfo  {journal} {JCAP}\ }\textbf {\bibinfo {volume} {05}},\
  \bibinfo {pages} {023} (\bibinfo {year} {2021})},\ \Eprint
  {http://arxiv.org/abs/2101.11244} {arXiv:2101.11244 [astro-ph.CO]}
  \BibitemShut {NoStop}%
\end{thebibliography}%

\end{document}